\documentclass[12pt,a4paper,oneside]{article}
\usepackage[centertags]{amsmath}
\usepackage{stmaryrd}
\usepackage{dsfont}
\usepackage{url}
\usepackage{geometry}
\usepackage{hyperref}
\usepackage[latin1]{inputenc}
\usepackage{amsfonts,amssymb,amsmath,amscd,slashed,amsthm}
\usepackage{enumerate}
\usepackage[dvips]{graphicx}
\usepackage{color}
\usepackage{verbatim}
\usepackage{psfrag}
\usepackage{epsfig}

\usepackage{epsfig}
\usepackage{epstopdf}

\usepackage{hyphenat}
\usepackage[english]{babel}

\newtheorem{Definition}{Definition}
\newtheorem{Conjecture}{Conjecture}
\newtheorem{Example}{Example}
\newtheorem{Theorem}{Theorem}
\newtheorem{Claim}{Claim}
\newtheorem{Lemma}{Lemma}
\newtheorem{Remark}{Remark}
\newtheorem{Corollary}{Corollary}
\newtheorem{Proposition}{Proposition}

\newenvironment{Proof}{\textbf{Proof}}{ \qed \\}
\newenvironment{Thm}[1][]{\begin{Theorem} \normalfont \textbf{#1} \itshape}{\end{Theorem}}

\newenvironment{Def}[1][]{\begin{Definition} \normalfont \textbf{#1} \itshape}{\end{Definition}}
\newenvironment{Ex}[1][]{\begin{Example} \normalfont \textbf{#1}}{\qed \end{Example}}
\newenvironment{Lem}[1][]{\begin{Lemma} \normalfont \textbf{#1} \itshape}{\end{Lemma}}
\newenvironment{Rem}[1][]{\begin{Remark} \normalfont \textbf{#1}}{\end{Remark}}
\newenvironment{Cor}[1][]{\begin{Corollary} \normalfont \textbf{#1}}{\end{Corollary}}
\newenvironment{Prop}[1][]{\begin{Proposition} \normalfont \textbf{#1}}{\end{Proposition}}

\newcommand{\supp}{\textnormal{supp}}
\newcommand{\pr}{\textnormal{pr}}

\newcounter{mnotecount}[section]
\renewcommand{\themnotecount}{\thesection.\arabic{mnotecount}}
\newcommand{\mnote}[1]
{\protect{\stepcounter{mnotecount}}$^{\mbox{\footnotesize
$
\bullet$\themnotecount}}$ \marginpar{
\raggedright\tiny\em
$\!\!\!\!\!\!\,\bullet$\themnotecount: #1} }

\definecolor{darkgreen}{rgb}{0,.5,0}

\newcommand{\A}{\mathcal{A}}
\newcommand{\B}{\mathcal{B}}

\newcommand{\F}{\mathcal{F}}

\newcommand{\M}{\mathcal{M}}

\newcommand{\K}{\mathcal{K}}
\newcommand{\cL}{\mathcal{L}}
\newcommand{\T}{\mathcal{T}}

\newcommand{\R}{\mathcal{R}}
\newcommand{\Sw}{\mathcal{S}}
\newcommand{\X}{\mathcal{X}}

\newcommand{\jdn}{\textbf{1}}

\renewcommand{\bar}{\overline}		

\renewcommand{\R}{\mathbb{R}}
\newcommand{\PP}{\mathfrak{P}}

\begin{document}

\title{Causality for nonlocal phenomena}
\author{
Micha\l\  Eckstein$^{a,b}$, Tomasz Miller$^{c,b}$
}

\date{{\footnotesize $^a$ Faculty of Physics, Astronomy and Applied Computer Science, Jagiellonian University}\\
{\footnotesize ul. prof. Stanis{\l}awa {\L}ojasiewicza 11, 30-348 Krak\'ow, Poland}\\[0.3cm]
{\footnotesize $^b$ Copernicus Center for Interdisciplinary Studies}\\
{\footnotesize ul. S{\l}awkowska 17, 31-016 Krak\'ow, Poland}\\[0.3cm]
{\footnotesize $^c$ Faculty of Mathematics and Information Science, Warsaw University of Technology}\\
{\footnotesize ul. Koszykowa 75, 00-662 Warsaw, Poland}\\[0.3cm]
{\footnotesize michal.eckstein@uj.edu.pl} \quad{\footnotesize T.Miller@mini.pw.edu.pl}
}

\maketitle

\begin{abstract}
Drawing from the~theory of optimal transport we propose a~rigorous notion of a~causal relation for Borel probability measures on a~given spacetime. To prepare the~ground, we explore the~borderland between causality, topology and measure theory. We provide various characterisations of the~proposed causal relation, which turn out to be equivalent if the~underlying spacetime has a~sufficiently robust causal structure. We also present the~notion of the~`Lorentz--Wasserstein distance' and study its basic properties. Finally, we discuss how various results on causality in quantum theory, aggregated around Hegerfeldt's theorem, fit into our framework.
\end{abstract}

\section{Introduction}

The~notion of a~space, understood as a~set of points, provides an~indispensable framework for every physical theory. But, regardless of the~physical system that is being modelled, the~space itself is not directly observable. Indeed, any measuring apparatus can provide information about the~localisation only up to a~finite resolution. In the~relativistic context, it means that the~\emph{event} is an~idealised concept, which is not accessible to any observer.

Apart from the~`practical' obstructions for measuring position, there exist also fundamental ones due to the~quantum effects manifest at small scales. Although non-relativistic quantum mechanics does not impose any \textit{a priori} restrictions on the~accuracy of the~position measurement, in quantum field theory a~suitable `position operator' is always nonlocal (see for instance \cite{Fredenhagen82,Foldy,Thaller}). Moreover, an~attempt to perform a~very accurate measurement of localisation in spacetime would require the~use of signals of very short wavelength, resulting in an~extreme concentration of energy. The~latter would eventually lead to black hole formation and the~desired information would become trapped \cite{DFR94,DFR95}.

It is generally believed that any physical theory should be causal, i.e. that no information can be transmitted with the~speed exceeding the~velocity of light. Indeed, despite some controversies (compare for example \cite{Nimtz} and \cite{Hegerfeldt1998}), no evidence of a~physical process that would involve superluminal signalling was found in any system (see for instance \cite{SuperluminalWinful}), even at the~level of Planck scale \cite{DSRFermi}.

In relativity theory it is straightforward to implement the~postulate of causality as the~Lorentzian metric induces a~precise notion of causal curves. Although Einstein's equations admit spacetime solutions with closed causal curves, they are usually discarded as unphysical \cite{Hawking}.


On the~other hand, the~status of causality in quantum theory is much more subtle because of its nonlocal nature. Hegerfeldt's theorem \cite{Hegerfeldt1} (see also \cite{Hegerfeldt1985,Hegerfeldt2001,Hegerfeldt2}) implies that during the~evolution of a~generic quantum system driven by a~Hamiltonian bounded from below, an~initially localised state\footnote{`Localised' in the~context of Hegerfeldt's theorem usually means of compact support in space, but the~argument extends to states with exponentially bounded tails \cite{Hegerfeldt1985}.} immediately develops infinite tails. However, whereas initial localisation implies the~breakdown of Einstein causality, the~use of nonlocal states does not guarantee a~subluminal evolution.  In fact, the~results of Hegerfeldt suggest \cite{Hegerfeldt1985} that acausal evolution is a~feature of the~quantum \emph{system} and not the~\emph{state}. In other words, if a~system impels a~superluminal propagation one could use nonlocal states to effectuate a~faster-than-light communication.


In quantum field theory the~nonlocality is even more prevailing, but it does not allow for communication between spacelike separated regions of spacetime \cite{CausalityQFT}. There is thus strong evidence that quantum theory, despite its inherent nonlocality, conforms to the~principle of causality \cite{QIandGR}. In fact, the~request of no faster-than-light signalling is often used as a~guiding principle to restrain the~admissible quantum theories \cite{CausalityOps,CausalityQM} and their possible extensions \cite{InformationCausality}. In quantum field theory it is reflected by promoting the~principle of microscopic causality to one of the~axioms \cite{Haag,Wightman}.

However, the~study of causation in quantum theory (and other nonlocal theories) is far from being complete. 
One of the~stumbling blocks is the~lack of a~suitable notion of causality for nonlocal objects, like wave functions. To properly investigate Einstein's principle, one needs to disentangle nonlocality from the~potential causality violation, as for instance the~interference fringes can travel superluminally, but cannot be utilised to send information \cite{Berry2012}. Also --- to our knowledge --- in the~study of causality in quantum systems, time was treated as an~external parameter, whereas the~most riveting consequences of Einstein causality, in particular the~existence of horizons, manifest themselves in curved spacetimes.


The~aim of this paper is to provide a~rigorous notion of a~causal relation between probability measures on a~given spacetime. These can be utilised to model classical spread objects, for instance charge or energy density, as well as quantum probabilities obtained via the~`modulus square principle' from wave functions. Moreover, one can make use of probability measures to take into account experimental errors, as the~measurement of any physical object's spacetime localisation would effectively be vitiated by an~error resulting from the~apparatus' imperfection.

We allow the~probability measures to be spread also in the~timelike direction, as typical states in quantum field theory extend over the~whole spacetime \cite{CausalityQFT2}. We work in a~generally covariant framework, hence our definitions and results apply to any curved spacetime with a~sufficiently rich causal structure.

The~paper is organised as follows: In Section \ref{sec:prelim} we recall some basic notions in topology, measure theory and causality, to make the~paper self-contained and accessible to a~broad range of researchers. Section \ref{sec:Borel} contains the~first result on the~verge of causality and measurability, which establishes the~foundations for the~developed theory. The~main concepts and results of the~paper are aggregated in Section \ref{sec:main}. We start off with a~`dual' definition of the~causal relation, based on the~notion of \emph{causal functions} \cite[Definition 2.3]{Moretti}, proposed in \cite{CQG2013} in a~much wider context of noncommutative geometry. In several steps we show that it encapsulates an~intuitive notion of causality for nonlocal objects:

\begin{quote}
Each infinitesimal part of the~probability density should travel along a~future-directed causal curve.
\end{quote}

At each step we keep the~causality conditions imposed on the~underlying spacetime as low as possible. At the~same time we provide several characterisations of causality for probability measures, which illustrate the~concept and provide tools for concrete computations.

Motivated by the~main result, we put forward in Section \ref{subsec:omega} a~definition of a~causal relation between the~probability measures, valid on any spacetime, and study its properties. In particular, we demonstrate that the~proposed relation is a~partial order in the~space of Borel probability measures on a~given spacetime $\M$, even with a~relatively poor causal structure.

Finally, drawing from the~theory of optimal transport adapted to the~relativistic setting we propose in Section \ref{sec:LW} a~notion of the~`Lorentz--Wasserstein' distance in the~space of measures.


We conclude, in Section \ref{sec:outlook}, with an~outlook into the~possible future developments and applications. In particular, we briefly discuss the potential use of the presented results in the study of causality in quantum theory. We also address the~interrelation of probability measures with states on $C^*$-algebras. In this way we provide a~link with the~notion of `causality in the~space of states' proposed originally in \cite{CQG2013} in the~framework of noncommutative geometry.

\section{Preliminaries}\label{sec:prelim}

Throughout the~paper we denote $\mathbb{N} := \{1,2,\ldots\}$ and $\mathbb{N}_0 := \mathbb{N} \cup \{0\}$.

The~space of continuous, continuous and bounded, continuous and compactly supported real-valued functions on a~topological space $\M$ will be respectively denoted by $C(\M)$, $C_b(\M)$, $C_c(\M)$. Analogous spaces of smooth functions will be respectively denoted by $C^\infty(\M)$, $C_b^\infty(\M)$, $C_c^\infty(\M)$.

\subsection{Topology}\label{subsec:topology}

Let $\M$ denote a~topological space and let $\X \subseteq \M$. The~closure, interior, boundary and complement of $\X$ will be denoted, respectively, by $\bar{\X}$, $\textnormal{int} \, \X$, $\partial \X$ and $\X^c$.

An~\emph{open cover} of $\X \subseteq \M$ is a~family $\{U_\alpha\}_{\alpha \in A}$ of open subsets of $\M$ such that $\bigcup\limits_{\alpha \in A} U_\alpha \supseteq \X$. $\M$ is called \emph{Lindel\"{o}f} iff every its open cover has a~countable subcover.

A~subset $\X \subseteq \M$ is called \emph{compact} iff every its open cover has a~finite subcover. It is called \emph{sequentially compact} iff every sequence in $\X$ has a~subsequence convergent in $\X$. It is called \emph{precompact} (or \emph{relatively compact}) iff its closure is compact. Finally, it is called \emph{$\sigma$-compact} iff it is a~countable union of compact subsets. In particular, $\M$ is $\sigma$-compact if and only if it admits an~\emph{exhaustion by compact sets}, that is a~sequence $(K_n)_{n \in \mathbb{N}}$ of compact sets such that $K_n \subseteq K_{n+1}$ and $\bigcup_{n=1}^\infty K_n = \M$.

If $\M$ is Hausdorff, then every its compact subset is closed. If $\M$ is \emph{second-countable}, that is if $\M$ has a~countable base, then the~notions of compactness and sequential compactness coincide.


A~Hausdorff\footnote{There exist many definitions of local compactness, which are all equivalent in Hausdorff spaces.} space $\M$ is called \emph{locally compact} iff every its point has a~precompact neighbourhood.

$\M$ is called \emph{separable} iff there exists a~countable subset $\{a_n\}_{n \in \mathbb{N}} \subseteq \M$ dense in $\M$.
Every open subspace of a~separable space is itself separable. $\M$ is called \emph{(completely) metrisable} iff there exists a~(complete) metric $\rho: \M^2 \rightarrow \mathbb{R}_{\geq 0}$ inducing its topology. Fixing a~metric allows one to talk about \emph{balls}. By $B(x,\varepsilon) := \{ y \in \M \ | \ \rho(x,y) < \varepsilon\}$ we denote an~\emph{open ball} centered at $x \in \M$ of radius $\varepsilon > 0$. By $\bar{B}(x,\varepsilon)$ we denote its closure. Finally, $\M$ is called \emph{Polish} iff it is separable and completely metrisable.

In the~following, we are going to work with \emph{spacetimes} (see Section \ref{subsec:causality}), which are examples of second-countable locally compact Hausdorff (LCH) spaces. Every such space is
\begin{itemize}
\item Lindel\"{o}f \cite[Theorem 16.9]{Willard};
\item Polish, because \cite[Theorems 19.3 \& 23.1]{Willard} imply that every second-countable LCH is metrisable, and by \cite[Corollary 2.3.32]{Srivastava} this means that every second-countable LCH is Polish;
\item $\sigma$-compact, because by taking a~countable set $\{a_n\}_{n \in \mathbb{N}}$ dense in $\M$ (existing by separability), and denoting by $U_n$ a~precompact neighbourhood of $a_n$ (existing by local compactness), one has that $\M = \bigcup\limits_{n = 1}^\infty \bar{U}_n$.
\end{itemize}

Moreover, every open subspace of a~second-countable LCH space is itself a~second-countable LCH space \cite[Theorem 18.4]{Willard}.

LCH spaces satisfy the~somewhat modified version of Urysohn's lemma \cite[2.12]{RudinRCA}
\begin{Thm}(Urysohn's lemma, LCH version)
\label{UrysohnLCH}
Let $\M$ be a~LCH space and let $\K \subseteq U \subseteq \M$, where $\K$ is compact and $U$ is open. Then, there exists $f \in C_c(\M)$ such that $f|_\K \equiv 1$, $0 \leq f \leq 1$ and $\supp \, f \subseteq U$.
\end{Thm}

\subsection{Measure theory}\label{subsec:measure}


Let $\M$ be a~topological space. The~\emph{$\sigma$-algebra of Borel sets} $\B(\M)$ is the~smallest family of subsets of $\M$ containing the~open sets, which is closed under complements and countable unions (and hence also under countable intersections). If $\M$ is a~Hausdorff space then, in particular, its $\sigma$-compact subsets are Borel.

A~function $f: \M_1 \rightarrow \M_2$ between topological spaces is called \emph{Borel} iff $f^{-1}(V) \in \B(\M_1)$ for any $V \in \B(\M_2)$. Every continuous (or even semi-continuous) real-valued function is Borel, but not \emph{vice versa}.

A~\emph{Borel probability measure} on $\M$ is a~function $\mu: \B(\M) \rightarrow \mathbb{R}_{\geq 0}$ satisfying $\mu(\M) = 1$ and $\mu\left( \bigcup\limits_{n=1}^\infty \X_n \right) = \sum\limits_{n=1}^\infty \mu(\X_n)$ for any $\{\X_n\}_{n \in \mathbb{N}} \subseteq \B(\M)$ such that $\X_n \cap \X_m = \emptyset$ for $n \neq m$. A~Borel set whose measure $\mu$ is zero is called \emph{$\mu$-null}. The~pair $(\M, \mu)$ is called a~\emph{probability space}. The~set of Borel probability measures on $\M$ will be denoted by $\mathfrak{P}(\M)$.

\pagebreak
Every $\mu \in \mathfrak{P}(\M)$ has the~following properties \cite[Theorem 1.19]{RudinRCA}:
\begin{itemize}
\item $\mu(\emptyset) = 0$;
\item $\mu$ is \emph{monotone}, i.e. $\forall \, \X_1,\X_2 \in \B(\M) \quad \X_1 \subseteq \X_2 \ \Rightarrow \ \mu(\X_1) \leq \mu(\X_2)$;
\item $\mu$ is \emph{countably subadditive}, i.e $\forall \, \{\X_n\}_{n \in \mathbb{N}} \subseteq \B(\M) \quad \mu\left( \bigcup\limits_{n=1}^\infty \X_n \right) \leq \sum\limits_{n=1}^\infty \mu(\X_n)$;
\item for any sequence $(\X_n)_{n \in \mathbb{N}} \subseteq \B(\M)$ which is \emph{increasing}, i.e. $\X_n \subseteq \X_{n+1}$, it is true that
\begin{align}
\label{meas1}
\mu\left( \bigcup\limits_{n=1}^\infty \X_n \right) = \lim\limits_{n \rightarrow +\infty} \mu(\X_n);
\end{align}
\item for any sequence $(\X_n)_{n \in \mathbb{N}} \subseteq \B(\M)$ which is \emph{decreasing}, i.e. $\X_n \supseteq \X_{n+1}$, it is true that
\begin{align}
\label{meas2}
\mu\left( \bigcap\limits_{n=1}^\infty \X_n \right) = \lim\limits_{n \rightarrow +\infty} \mu(\X_n).
\end{align}
\end{itemize}
Furthermore, if $\M$ is metrisable, then every $\mu \in \mathfrak{P}(\M)$ is \emph{regular} \cite[Lemma 3.4.14]{Srivastava}, i.e.
\begin{align}
\label{meas3}
\forall \, \X \in \B(\M) \quad \mu(\X) & = \sup \left\{ \mu(F) \ | \ F \subseteq \X, \, F \textnormal{ closed} \right\}
\\
\nonumber
& = \inf \left\{ \mu(U) \ | \ U \supseteq \X, \, U \textnormal{ open} \right\}.
\end{align}
Finally, if $\M$ is Polish, then every $\mu \in \mathfrak{P}(\M)$ is also \emph{tight} \cite[Theorem 3.4.20]{Srivastava}, i.e.
\begin{align}
\label{meas4}
\forall \, \X \in \B(\M) \quad \mu(\X) = \sup \left\{ \mu(\K) \ | \ \K \subseteq \X, \, \K \textnormal{ compact} \right\}
\end{align}
Borel probability measures with properties \eqref{meas3} and \eqref{meas4} are called \emph{Radon probability measures}. Since we will be working with spacetimes (which are Polish spaces), all elements of $\mathfrak{P}(\M)$ will be Radon. For simplicity, from now on the~term `measure' will always stand for the~`Borel probability measure'.

For any $\X \subseteq \M$ its \emph{indicator function}\footnote{The indicator function $\jdn_\X$ is sometimes called the \emph{characteristic function of $\X$}, but this term has another unrelated meaning in probability theory, which might cause confusion.} $\jdn_\X : \M \rightarrow \mathbb{R}$ is defined by $\jdn_\X(p) = 1$ for $p \in \X$ and $\jdn_\X(p) = 0$ otherwise. $\jdn_\X$ is a~Borel function iff $\X \in \B(\M)$.

A~\emph{simple function} on $\M$ is any function $s: \M \rightarrow \mathbb{R}$ whose range $s(\M)$ is finite. Such a~function can be written in the~form $s = \sum\limits_{i=1}^{n} \alpha_i \jdn_{\X_i}$ where $s(\M) = \{\alpha_1,\ldots,\alpha_n\}$ and $\X_i = s^{-1}(\alpha_i)$ ($i=1,\ldots,n$). Notice that $s$ is Borel iff all $\X_i$'s are Borel sets.

For any $\mu \in \mathfrak{P}(\M)$ the~(Lebesgue) \emph{integral} of a~Borel nonnegative function $f$ is defined as
\begin{align*}
\int\limits_\M f d\mu := \sup \left\{ \sum\limits_{i=1}^{n} \alpha_i \mu(\X_i) \ \big| \ \sum\limits_{i=1}^{n} \alpha_i \jdn_{\X_i} \leq f \right\}.
\end{align*}
It is well-defined by \cite[Theorem 1.17]{RudinRCA}, albeit it might be infinite. Now for any Borel function $f$ introduce two nonnegative Borel functions $f^\pm := \max\{ \pm f, 0 \}$ and define $\int\limits_\M f d\mu := \int\limits_\M f^+ d\mu - \int\limits_\M f^- d\mu$ if at least one of the~integrals is finite. For any $\X \in \B(\M)$ one additionally defines $\int\limits_\X f d\mu := \int\limits_\M \jdn_\X f d\mu$. A~function $f$ is \emph{$\mu$-integrable} iff it is Borel and $\int\limits_\M |f| d\mu < +\infty$. The~space of $\mu$-integrable functions is denoted by $\cL^1(\M,\mu)$. Observe that Borel bounded functions are $\mu$-integrable for any $\mu \in \mathfrak{P}(\M)$.

We will often use the~following classical theorem \cite[Theorem 1.34]{RudinRCA}.
\begin{Thm}(Lebesgue's dominated convergence theorem)
Let $(f_n)_{n \in \mathbb{N}}$ be a~sequence of Borel functions on $\M$ such that $f_n \rightarrow f$ pointwise. For any $\mu \in \mathfrak{P}(\M)$, if there exists $g \in \cL^1(\M, \mu)$ such that $|f_n| \leq g$ for all $n \in \mathbb{N}$, then also $f \in \cL^1(\M, \mu)$ and
\begin{align*}
\lim\limits_{n \rightarrow +\infty} \int\limits_\M f_n d\mu = \int\limits_\M f d\mu.
\end{align*}
\end{Thm}
We also recall another classical result, which allows to define a~Radon probability measure on a~LCH space $\M$ by means of a~positive linear functional on $C_c(\M)$ of norm 1 \cite[Theorem 2.14]{RudinRCA}.
\begin{Thm}(Riesz--Markov--Kakutani representation theorem)
\label{Riesz}
Let $\M$ be a~LCH space and let $\Lambda: C_c(\M) \rightarrow \mathbb{R}$ be a~linear map such that
\begin{itemize}
\item $\Lambda (f) \geq 0$ for all nonnegative $f \in C_c(\M)$,
\item $\sup\limits_{\|f\|=1} |\Lambda (f)| = 1$, where $\| \cdot \|$ denotes the~supremum norm.
\end{itemize}
Then, there exists a~unique Radon probability measure $\mu \in \mathfrak{P}(\M)$ such that $\Lambda (f) = \int\limits_\M f d\mu$ for all $f \in C_c(\M)$.
\end{Thm}

Any Borel function $f: \M_1 \rightarrow \M_2$ between topological spaces induces the~\emph{pushforward map} $f_\ast: \mathfrak{P}(\M_1) \rightarrow \mathfrak{P}(\M_2)$, $\mu \mapsto f_\ast\mu$. The~latter is called a~\emph{pushforward measure} and is defined through
\begin{align*}
\forall \, V \in \B(\M_2) \quad f_\ast\mu(V) := \mu\left(f^{-1}(V)\right).
\end{align*}
As for the~integrability, one has that $g \in \cL(\M_2, f_\ast\mu)$ iff $g \circ f \in \cL(\M_1, \mu)$, in which case $\int\limits_{\M_2} g \, d(f_\ast\mu) = \int\limits_{\M_1} g \circ f \, d\mu$.

Given two probability spaces $(\M_1, \mu_1), (\M_2, \mu_2)$, there exists a~unique measure $\mu_1 \times \mu_2 \in \mathfrak{P}(\M_1 \times \M_2)$, called the~\emph{product measure}, such that $(\mu_1 \times \mu_2)(U_1 \times U_2) = \mu_1(U_1)\mu_2(U_2)$ for any $U_i \in \B(\M_i)$, $i=1,2$ (cf. \cite[Chapter 7]{RudinRCA} for details).

On the~other hand, given $\omega \in \mathfrak{P}(\M_1 \times \M_2)$, its \emph{marginals} are defined as $(\pr_i)_\ast \omega \in \mathfrak{P}(\M_i)$, where $\pr_i: \M_1 \times \M_2 \rightarrow \M_i$ ($i=1,2$) are the~canonical projection maps. Obviously, the~marginals of the~product measure $\mu_1 \times \mu_2$ are $\mu_1$ and $\mu_2$, however, usually there are many measures on $\M_1 \times \M_2$ sharing the~same pair of marginals.

Given a~measure $\mu \in \mathfrak{P}(\M)$, its \emph{support} can be defined as the~smallest closed set with full measure. Symbolically, $\supp \, \mu := \bigcap \left\{ F \subseteq \M : F \textnormal{ closed, } \mu(F) = 1 \right\}$.


\subsection{Causality theory}\label{subsec:causality}

For a~detailed exposition of causality theory the~reader is referred to \cite{Beem,MS08,BN83,Penrose1972}.

Recall that a~\emph{spacetime} is a~connected time-oriented Lorentzian manifold. Causality theory introduces and studies certain binary relations between points (i.e. \emph{events}) of a~given spacetime $\M$. Namely, for any $p,q \in \M$, we say that $p$ \emph{causally (chronologically, horismotically) precedes} $q$, what is denoted by $p \preceq q$ (resp. $p \ll q$, $p \rightarrow q$), iff there exists a~piecewise smooth future-directed causal (resp. timelike, null) curve $\gamma: [0,1] \rightarrow \M$ \emph{from $p$ to $q$}, i.e. $\gamma(0) = p$ and $\gamma(1) = q$.

Clearly the~relations $\preceq$ and $\ll$ are transitive and $\preceq$ is also reflexive. Moreover (\cite[Chapter 14, Corollary 1]{BN83}),
\begin{align}
\label{prelim1}
\forall \, p,q,r \in \M \quad p \ll r \preceq q \ \vee \ p \preceq r \ll q \ \ \Rightarrow \ \ p \ll q.
\end{align}

To denote $\preceq$ ($\ll$, $\rightarrow$) understood as a~subset of $\M^2$ it is customary to use the~symbol $J^+$ (resp. $I^+$, $E^+$). $I^+$ is open and equal to $\textnormal{int} \, J^+$, and so the~\emph{causal structure} of $\M$ is completely determined by the~relation $\preceq$ and the~topology of $\M$. Moreover, $\bar{I}^+ = \bar{J}^+$, $\partial I^+ = \partial J^+$ and $E^+ = J^+ \setminus I^+$.

For any $\X \subseteq \M$ one defines
\begin{align}
\label{fpsets1}
J^+(\X) := \pr_2\left( (\X \times \M) \cap J^+ \right) \quad \textnormal{and} \quad J^-(\X) := \pr_1\left( (\M \times \X) \cap J^+ \right).
\end{align}
If $\X$ is a~singleton, one simply writes $J^\pm(p)$ instead of $J^\pm(\{p\})$. Notice that $J^\pm(\X) = \bigcup\limits_{p \in \X} J^\pm(p)$.

Let now $U \subseteq \M$ be an~open subset of $\M$. One defines $\preceq_U$ to be the~causal precedence relation on $U$ treated as a~spacetime on its own right. By analogy with $J^+$, we denote $J^+_U := \{ (p,q) \in U^2 \ | \ p \preceq_U q \}$. Notice that $J^+_U \subseteq J^+ \cap U^2$, but not necessarily \emph{vice versa} because $p \preceq_U q$ requires a~piecewise smooth future-directed causal curve from $p$ to $q$ not only to exist, but also to be contained in $U$.

Analogously to (\ref{fpsets1}), one defines $J^\pm_U(\X)$ for any subset $\X \subseteq \M$.

One similarly introduces $I^\pm(\X), I^+_U, I^\pm_U(\X)$ as well as $E^\pm(\X), E^+_U, E^\pm_U(\X)$. Observe that, by (\ref{prelim1}), $J^+(\X) = I^+(\X)$ for any open $\X \subseteq \M$.

A~subset ${\cal F} \subseteq \M$ is called a~\emph{future set} iff\footnote{Notice that only the~inclusion `$\subseteq$' is nontrivial in the~definition of a~ future (past) set.} $J^+({\cal F}) = {\cal F}$. Similarly, subset ${\cal P} \subseteq \M$ is called a~\emph{past set} iff $J^-({\cal P}) = {\cal P}$. Usually it is required that future and past sets be open by definition. However, if we drop this assumption future and past sets behave more naturally under set-theoretical operations.
\begin{Prop}
\label{prelimLem1}
$\F \subseteq \M$ is a~future set iff $\F^c$ is a~past set.
\end{Prop}
\begin{Proof}\textbf{:}
The~statement is proven by the~following chain of equivalences:
\begin{align*}
J^-(\F^c) \subseteq \F^c \quad & \Leftrightarrow \quad \forall \, s \in \M \ \left[ \exists \, r \in \F^c \ s \preceq r \right] \ \Rightarrow \ s \in \F^c \\
& \Leftrightarrow \quad \forall \, s \in \M \ \left[ \exists \, r \in J^+(s) \setminus \F \right] \ \Rightarrow \ s \not\in \F \\
& \Leftrightarrow \quad \forall \, s \in \M \ \ J^+(s) \setminus \F = \emptyset \ \Leftarrow \ s \in \F \\
& \Leftrightarrow \quad \forall \, s \in \F \ \ J^+(s) \subseteq \F \quad \Leftrightarrow \quad \bigcup\limits_{s \in \F} J^+(s) \subseteq \F \quad \Leftrightarrow \quad J^+(\F) \subseteq \F.
\end{align*}
\end{Proof}
\begin{Prop}
\label{prelimLem2}
Let $\{ \X_\alpha \}_{\alpha \in A}$ be a~family of future (past) subsets of $\M$. Then also $\bigcup\limits_{\alpha \in A} \X_\alpha$ and $\bigcap\limits_{\alpha \in A} \X_\alpha$ are future (past) subsets of $\M$.
\end{Prop}
\begin{Proof}\textbf{:}
Assuming that all $\X_\alpha$'s are future sets, notice that $J^+\left( \bigcup\limits_{\alpha \in A} \X_\alpha \right) = \bigcup\limits_{\alpha \in A} J^+\left( \X_\alpha \right) = \bigcup\limits_{\alpha \in A} \X_\alpha$. If $\X_\alpha$'s are past sets, simply replace $J^+$ with $J^-$ in the~previous sentence.

We have thus shown that a~union of the~family of future (past) sets is a~future (past) set. To obtain an~analogous result for the~intersection, one simply uses Proposition \ref{prelimLem1} and de Morgan's laws.
\end{Proof}

A~function $f: \M \rightarrow \mathbb{R}$ is called
\begin{itemize}
\item a~\emph{causal function} iff it is non-decreasing along every future-directed causal curve;
\item a~\emph{generalised time function} iff it is increasing along every future-directed causal curve;
\item a~\emph{time function} iff it is a~continuous generalised time function;
\item a~\emph{temporal function} iff it is a~smooth function with past-directed timelike gradient.
\end{itemize}
Each of the above properties is stronger than the~preceding one.

Causal functions can be characterised by means of future sets.
\begin{Prop}
\label{prelimLem3}
Let $\M$ be a~spacetime. For any function $f: \M \rightarrow \mathbb{R}$ the~following conditions are equivalent
\begin{enumerate}[i)]
\item $f$ is causal,
\item $f^{-1}((a,+\infty))$ is a~future set for any $a \in \mathbb{R}$,
\item $f^{-1}([a,+\infty))$ is a~future set for any $a \in \mathbb{R}$.
\end{enumerate}
\end{Prop}
\begin{Proof}\textbf{:}
`i) $\Rightarrow$ ii)' Assume that $f$ is causal and $a \in \mathbb{R}$. If $f^{-1}((a,+\infty)) = \emptyset$, then it is trivially a~future set. Suppose then that $f^{-1}((a,+\infty)) \neq \emptyset$ and take any $q \in J^+\left(f^{-1}((a,+\infty))\right)$, which means that there exists $p \preceq q$ such that $f(p) > a$. By causality of $f$ we have that $f(q) \geq f(p) > a$ and so $q \in f^{-1}((a,+\infty))$. We thus obtain the~inclusion $J^+\left(f^{-1}((a,+\infty))\right) \subseteq f^{-1}((a,+\infty))$. The~other inclusion is obvious.

`ii) $\Rightarrow$ iii)' Observe that $f^{-1}([a,+\infty)) = \bigcap_{n=1}^\infty f^{-1}((a-\tfrac{1}{n},+\infty))$. By ii) and Proposition \ref{prelimLem2}, we obtain iii).

`iii) $\Rightarrow$ i)' Assume $f$ is not causal, i.e. there exist $p,q \in \M$ such that $p \preceq q$ but $f(p) > f(q)$. We claim that $f^{-1}\left([f(p),+\infty)\right)$ is not a~future set. Indeed, were it a~future set, then, since it clearly contains $p$, it would contain $q$ as well. But this would mean that $f(q) \geq f(p)$, in contradiction with the assumption.
\end{Proof}

On the~other hand, future sets can be characterised by means of their indicator function being causal.
\begin{Cor}
\label{futurepastsets}
Let $\M$ be a~spacetime. $\F \subseteq \M$ is a~future set iff the~function $\jdn_\F$ is causal.
\end{Cor}
\begin{Proof}\textbf{:}
Observe that
\begin{align*}
\jdn_\F^{-1}\left( [a,+\infty) \right) = \left\{ \begin{array}{ll}
                                                        \M & \textnormal{for } a \leq 0 \\
                                                        \F & \textnormal{for } 0 < a \leq 1 \\
                                                        \emptyset & \textnormal{for } a > 1
                                                        \end{array} \right. \cdot
\end{align*}
By equivalence `i) $\Leftrightarrow$ iii)' from Proposition \ref{prelimLem3}, we immediately obtain the~desired equivalence.
\end{Proof}

An~\emph{admissible measure} on $\M$ is any $\eta \in \mathfrak{P}(\M)$ such that (\cite[Definiton 3.19]{Beem})
\begin{itemize}
\item for any nonempty open subset $U \subseteq \M$ $\ \eta(U) > 0$,
\item for any $p \in \M$ the~boundaries $\partial I^\pm(p)$ are $\eta$-null.
\end{itemize}
To such $\eta$ one associates the~functions $t^-, t^+: \M \rightarrow \mathbb{R}$, called \emph{past} and \emph{future} volume functions, respectively, defined via
\begin{align*}
\forall \, p \in \M \quad t^\pm(p) := \mp \eta(I^\pm(p)).
\end{align*}
Volume functions are causal and semi-continuous and hence Borel.

For any $p,q \in \M$ let $\hat{C}(p,q)$ denote the~set of piecewise smooth future-directed causal curves from $p$ to $q$. The~\emph{Lorentzian distance} (or \emph{time separation}) is the~map $d: \M^2 \rightarrow [0,+\infty]$ defined by
\begin{align*}
d(p, q) := \left\{ \begin{array}{ll}
                         \sup\limits_{\gamma \in \hat{C}(p,q)} \int\limits_0^1 \sqrt{-g_{\alpha\beta} \dot{\gamma}^\alpha \dot{\gamma}^\beta} dt & \textnormal{if } \hat{C}(p,q) \neq \emptyset
                         \\
                         0 & \textnormal{if } \hat{C}(p,q) = \emptyset
                         \end{array} \right. \cdot
\end{align*}
Its basic properties include:
\begin{enumerate}[i)]
\item For any $p,q \in \M$ $\ d(p,q) > 0 \quad \Leftrightarrow \quad p \ll q$.
\item The~\emph{reverse triangle inequality} holds. Namely, for any $p,q,r \in \M$
\begin{align}
\label{Ltriangle}
p \preceq r \preceq q \quad \Rightarrow \quad d(p, r) + d(r, q) \leq d(p, q).
\end{align}
\item If there exists a~timelike loop through $p \in \M$ (i.e. a~piecewise smooth curve from $p$ to $p$), then $d(p,p) = +\infty$. Otherwise $d(p,p) = 0$.
\item For any $p,q \in \M$, if $d(p, q) \in (0,+\infty)$ then $d(q, p) = 0$.
\item The~map $d$ is lower semi-continuous \cite[Chapter 14, Lemma 17]{BN83} and hence Borel.
\end{enumerate}

The~\emph{causal ladder} is a~hierarchy of spacetimes according to strictly increasing requirements on their causal properties \cite{Beem}. The~rungs of this ladder, from the~top to the~bottom, read:
\begin{align*}
& \textnormal{Globally hyperbolic} \ \; \Rightarrow \ \; \textnormal{Causally simple} \ \; \Rightarrow \ \; \textnormal{Causally continuous} \ \; \Rightarrow \ \; \textnormal{Stably causal}
\\
& \ \; \Rightarrow \ \; \textnormal{Strongly causal} \ \; \Rightarrow \ \; \textnormal{Distinguishing} \ \; \Rightarrow \ \; \textnormal{Causal} \ \; \Rightarrow \ \; \textnormal{Chronological}
\end{align*}
Each level of the~hierarchy can be defined in many equivalent ways. Below we present only these definitions, characterisations and properties, of which we make use in the~paper. For the~complete review of the~causal hierarchy, consult \cite[Section 3]{MS08}.
\pagebreak

$\M$ is \emph{chronological} iff it satisfies one of the~following equivalent conditions:
\begin{enumerate}[i)]
\item $p \not\ll p$ for all $p \in \M$.
\item No timelike loop exists.
\item Any volume function is increasing along every future-directed timelike curve.
\item $d(p,p) = 0$ for all $p \in \M$.
\end{enumerate}

$\M$ is \emph{causal} iff it satisfies one of the~following equivalent conditions:
\begin{enumerate}[i)]
\item The~relation $\preceq$ is a~\emph{partial order}, meaning that in addition to being reflexive and transitive, it is also antisymmetric.
\item No causal loop exists.
\end{enumerate}

$\M$ is \emph{future (past) distinguishing} iff it satisfies one of the~following equivalent conditions:
\begin{enumerate}[i)]
\item For any $p,q \in \M$, the~equality $I^+(p) = I^+(q)$ (resp. $I^-(p) = I^-(q)$) implies that $p=q$.
\item Any future (past) volume function is a~generalised time function \cite[Proposition 3.24]{Beem}.
\end{enumerate}
$\M$ is \emph{distinguishing} iff it is both future and past distinguishing.

$\M$ is \emph{strongly causal} iff the~family $\{I^+(p) \cap I^-(q) \ | \ p,q \in \M\}$ is a~base of the~standard manifold topology of $\M$. It is \emph{stably causal} iff it admits a~time function or, equivalently, iff it admits a~temporal function \cite{BS04}. It is \emph{causally continuous} iff any volume function is a~time function.

$\M$ is \emph{causally simple} iff it is causal and satisfies one of the~following equivalent conditions \cite[Proposition 3.68]{MS08}:
\begin{enumerate}[i)]
\item $J^+(p)$ and $J^-(p)$ are closed for every $p \in \M$;
\item $J^+(\K)$ and $J^-(\K)$ are closed for every compact $\K \subseteq \M$;
\item $J^+$ is a~closed subset of $\M^2$.
\end{enumerate}

Before providing a~definition of the~top level of the~causal hierarchy, recall that a~curve $\gamma: (a,b) \rightarrow \M$ with $-\infty \leq a < b \leq +\infty$ is called \emph{extendible} iff it has a~continuous extension onto $[a,b)$ or onto $(a,b]$. Otherwise such a~curve is called \emph{inextendible}. Recall also that a~\emph{Cauchy hypersurface} is a~subset $\Sw \subseteq \M$ which is met exactly once by any inextendible timelike curve. Any such $\Sw$ is a~closed \emph{achronal} (i.e. $\Sw^2 \cap I^+ = \emptyset$) topological hypersurface, met by every inextendible causal curve \cite[Chapter 14, Lemma 29.]{BN83}. However, such an $\Sw$ need not be \emph{acausal} (i.e. $\Sw^2 \cap J^+$ might be nonempty).

$\M$ is \emph{globally hyperbolic} iff it satisfies one of the~following equivalent conditions:
\begin{enumerate}[i)]
\item $\M$ is causal and the~sets $J^+(p) \cap J^-(q)$ are compact for all $p,q \in \M$;
\item $\M$ admits a~smooth temporal function $\T$, the~level sets of which are (smooth spacelike) Cauchy hypersurfaces \cite{BS04}.
\end{enumerate}
In a~globally hyperbolic spacetime the~Lorentzian distance $d$ is finite-valued and continuous. Moreover, for every $(p,q) \in J^+$ there exists a~causal geodesic $\gamma$ of length $d(p,q)$ \cite[Chapter 14]{BN83}.

\section{On the~$\sigma$-compactness of $J^+$}\label{sec:Borel}

The~purpose of this section is to prove the~following theorem.
\begin{Thm}
\label{mainThm}
Let $\M$ be a~spacetime. Then $J^+ \subseteq \M^2$ is a~$\sigma$-compact set.
\end{Thm}
Let us note here that property is automatic in causally simple spacetimes. Indeed, let $(K_n)_{n \in \mathbb{N}}$ be an~exhaustion of $\M$ with compact sets and notice that $J^+ = \bigcup\limits_{m,n \in \mathbb{N}} J^+ \cap (K_n \times K_m)$. But $J^+ \subseteq \M^2$ is a~\emph{closed} subset for $\M$ causally simple, therefore $J^+ \cap (K_n \times K_m)$ is compact for any $m,n \in \mathbb{N}$.

In the~proof of Theorem \ref{mainThm}, however, we shall make no assumptions on the~causal properties of $\M$.

Theorem \ref{mainThm} implies that $J^+$ is Borel for any spacetime. As we shall see, it also implies that $J^\pm(\X)$ is Borel for any closed $\X \subseteq \M$. Moreover, previous statements are still true if we replace $J^\pm$ with $E^\pm$.

Theorem \ref{mainThm} thus settled in the~overlap of causality theory, topology and measure theory. Whereas the~interplay between the~causal and topological properties of spacetimes is relatively well understood, the~question of Borelness of causal futures --- a~fundamental one from the~point of view of any conceivable measure-theoretical extension of causality theory --- has never been addressed to authors' best knowledge.
\\

We recall the~notion of \emph{simple convex sets} (called also \emph{simple regions}) \cite[Section 1]{Penrose1972}. Loosely speaking, they are small patches of the spacetime $\M$ with `nice' topological, differential and causal properties, and which constitute a countable cover of the~entire spacetime.

Concretely, let $\M$ be a~spacetime. Then for any $p \in \M$ there exists a~star-shaped neighbourhood $Q \subseteq T_p \M$ containing the~zero vector and such that the~exponential map $\exp_p$ restricted to $Q$ is a~diffeomorphism. The~image of this diffeomorphism $\exp_p(Q)$ is called a~\emph{normal neighbourhood of $p$}. Every event has a~neighborhood $U$ which is a~normal neighbourhood of any $p \in U$. Such $U$ is called \emph{convex}. If $U \subseteq \M$ is convex, then it is open and for any $p,q \in U$ there exists precisely one geodesic from $p$ to $q$ which is contained in $U$ \cite[p. 129]{BN83}.

From the~point of view of causality theory, the~following property of convex sets will be crucial: if $U \subseteq \M$ is convex, then $J^+_U$ is a~closed subset of $U^2$ \cite[Lemma 14.2]{BN83}.

Finally, a~convex set $N$ is called \emph{simple} iff it is precompact and contained in another convex set $U$.

Any spacetime $\M$ can be covered with a~family of simple convex sets \cite[Proposition 1.13]{Penrose1972}. This cover can be chosen countable, because every spacetime is a~Lindel\"{o}f space.

\begin{Proof}\textbf{ of Theorem \ref{mainThm}:}
Fix a~countable, locally finite family of simple convex sets $\{N_i\}_{i \in \mathbb{N}}$ covering $\M$. Let also $\{U_i\}_{i \in \mathbb{N}}$ be a family of convex sets such that $\forall \, i \in \mathbb{N} \ \bar{N}_i \subseteq U_i$, which exists by the~very definition of a~simple convex sets.

We introduce a~couple more definitions.

Take any $i \in \mathbb{N}$. Recall that $J^+_{U_i}$ is a~closed subset of $U_i^2$, whereas $\bar{N}_i^2$ is a~compact subset of $U_i^2$. Let us first define the~following compact subset of $U_i^2$
\begin{align*}
J^+_{(\bar{N}_i)} := J^+_{U_i} \cap \bar{N}_i^2 = \left\{ (p,q) \in \bar{N}_i^2 \ | \ p \preceq_{U_i} q \right\},
\end{align*}
\noindent
that is the~set containing all these pairs of points from $\bar{N}_i$ which can be connected by a~piecewise smooth future-directed causal curve contained in $U_i$. For any $\X \subseteq \M$ define, by analogy with (\ref{fpsets1}),
\begin{align}
\label{fpsets3}
J^+_{(\bar{N}_i)}(\X) := \pr_2\left( (\X \times \M) \cap J^+_{(\bar{N}_i)} \right) \quad \textnormal{and} \quad J^-_{(\bar{N}_i)}(\X) := \pr_1\left( (\M \times \X) \cap J^+_{(\bar{N}_i)} \right).
\end{align}
\noindent
If $\X$ is a~singleton, one writes simply $J^\pm_{(\bar{N}_i)}(p)$ instead of $J^\pm_{(\bar{N}_i)}(\{p\})$. Notice that if $\X$ is closed, then $J^\pm_{(\bar{N}_i)}(\X)$ is a~compact subset of $U_i$.

For the~next definition, fix $i_1, i_2 \in \mathbb{N}$ and introduce
\begin{align*}
J^+_{(\bar{N}_{i_1}, \bar{N}_{i_2})} & := \left\{ (p,q) \in \bar{N}_{i_1} \times \bar{N}_{i_2} \ | \ \exists \, r \in \bar{N}_{i_1} \cap \bar{N}_{i_2} \quad p \preceq_{U_{i_1}} r \preceq_{U_{i_2}} q \right\}
\\
& = \bigcup\limits_{r \in \bar{N}_{i_1} \cap \bar{N}_{i_2}} J^-_{\bar{N}_{i_1}}(r) \times J^+_{\bar{N}_{i_2}}(r).
\end{align*}
\noindent
This is the~set of all those pairs of points $(p,q) \in \bar{N}_{i_1} \times \bar{N}_{i_2}$, which can be connected by a~concatenation of two piecewise smooth future-directed causal curves, first of which is contained in $U_{i_1}$, while the~other in $U_{i_2}$, and the~concatenation point $r$ must lie in the~compact set $\bar{N}_{i_1} \cap \bar{N}_{i_2}$. As above, we additionally define, for any $\X \subseteq \M$,
\begin{align}
\label{fpsets4}
& J^+_{(\bar{N}_{i_1}, \bar{N}_{i_2})}(\X) := \pr_2\left( (\X \times \M) \cap J^+_{(\bar{N}_{i_1}, \bar{N}_{i_2})} \right) \quad \textnormal{and}
\\
\nonumber
& J^-_{(\bar{N}_{i_1}, \bar{N}_{i_2})}(\X) := \pr_1\left( (\M \times \X) \cap J^+_{(\bar{N}_{i_1}, \bar{N}_{i_2})} \right) \; .
\end{align}

Finally, fix $n \geq 3$ together with $i_1, i_2, \ldots, i_n \in \mathbb{N}$ and define, recursively,
\begin{align*}
& J^+_{(\bar{N}_{i_1}, \bar{N}_{i_2}, \ldots, \bar{N}_{i_n})}
\\
& := \left\{ (p,q) \in \bar{N}_{i_1} \times \bar{N}_{i_n} \ | \ \exists \, r \in \bar{N}_{i_{n-1}} \ (p, r) \in J^+_{(\bar{N}_{i_1}, \bar{N}_{i_2}, \ldots, \bar{N}_{i_{n-1}})} \ \wedge \ (r, q) \in J^+_{(\bar{N}_{i_{n-1}}, \bar{N}_{i_n})} \right\}
\\
& = \bigcup\limits_{r \in \bar{N}_{i_{n-1}}} J^-_{(\bar{N}_{i_1}, \bar{N}_{i_2}, \ldots, \bar{N}_{i_{n-1}})}(r) \times J^+_{(\bar{N}_{i_{n-1}}, \bar{N}_{i_n})}(r),
\end{align*}
where, for any $\X \subseteq \M$,
\begin{align}
\label{fpsets5}
& J^+_{(\bar{N}_{i_1}, \bar{N}_{i_2}, \ldots, \bar{N}_{i_n})}(\X) := \pr_2\left( (\X \times \M) \cap J^+_{(\bar{N}_{i_1}, \bar{N}_{i_2}, \ldots, \bar{N}_{i_n})} \right) \quad \textnormal{and}
\\
\nonumber
& J^+_{(\bar{N}_{i_1}, \bar{N}_{i_2}, \ldots, \bar{N}_{i_n})}(\X) := \pr_1\left( (\M \times \X) \cap J^+_{(\bar{N}_{i_1}, \bar{N}_{i_2}, \ldots, \bar{N}_{i_n})} \right) \; .
\end{align}

It is crucial to understand what these sets contain (see Figure \ref{Jp100}). Namely, $J^+_{(\bar{N}_{i_1}, \bar{N}_{i_2}, \ldots, \bar{N}_{i_n})}$ is the~set of all those pairs of points $(p,q) \in \bar{N}_{i_1} \times \bar{N}_{i_n}$ which can be connected by a~concatenation of $n-1$ piecewise smooth future-directed causal curves, each being of the~type discussed after the~definition of $J^+_{(\bar{N}_{i_1}, \bar{N}_{i_2})}$. The~curves' concatenation points must lie in $\bar{N}_{i_2}, \bar{N}_{i_3}, \ldots, \bar{N}_{i_{n-1}}$ (in that order).

\begin{figure}[h]
\begin{center}
\psfrag{p}{\footnotesize $p$}
\psfrag{q}{\footnotesize$q$}
\psfrag{n1}{\footnotesize$\!\!\bar{N}_{i_1}$}
\psfrag{n2}{\footnotesize$\bar{N}_{i_2}$}
\psfrag{n3}{\footnotesize$\bar{N}_{i_3}$}
\psfrag{n4}{\footnotesize$\bar{N}_{i_4}$}
\psfrag{u1}{\footnotesize$\!\!U_{i_1}$}
\psfrag{u2}{\footnotesize$U_{i_2}$}
\psfrag{u3}{\footnotesize$U_{i_3}$}
\psfrag{u4}{\footnotesize$U_{i_4}$}
\resizebox{!}{150pt}{\includegraphics[scale=0.6]{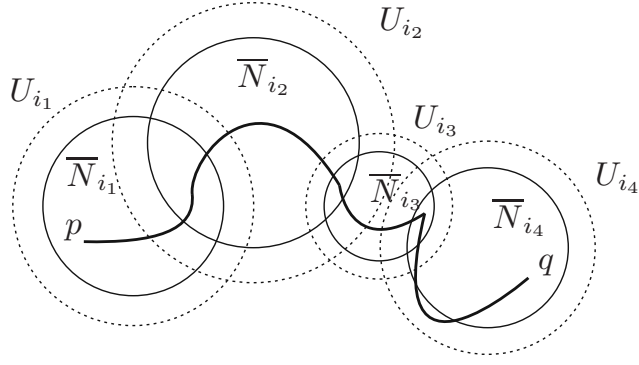}}
\caption{Here $(p,q) \in J^+_{(\bar{N}_{i_1}, \bar{N}_{i_2}, \bar{N}_{i_3}, \bar{N}_{i_4})}$. The piecewise smooth curve from $p$ to $q$ shown is assumed causal and future-directed.\label{Jp100}}
\end{center}
\end{figure}

We now claim and shall prove inductively that
\begin{align*}
& \forall \, n \geq 2 \ \ \forall i_1, i_2, \ldots, i_n \in \mathbb{N}
\\
& \qquad J^+_{(\bar{N}_{i_1}, \bar{N}_{i_2}, \ldots, \bar{N}_{i_n})} \textnormal{ is a~compact subset of } U_{i_1} \times U_{i_n}, \textnormal{ and hence of } \M^2.
\end{align*}
Let us first prove the~base case $n=2$. Let $\{a_m\}_{m \in \mathbb{N}}$ be a~dense subset of $N_{i_1} \cap N_{i_2}$, which exists by separability of $N_{i_1} \cap N_{i_2}$. Of course, $\{a_m\}_{m \in \mathbb{N}}$ is also a~dense subset of $\overline{N_{i_1} \cap N_{i_2}} = \bar{N}_{i_1} \cap \bar{N}_{i_2}$. Therefore, the~family of open balls $\{ B(a_m, \tfrac{1}{k}) \}_{m \in \mathbb{N}}$ is an~open cover of $\bar{N}_{i_1} \cap \bar{N}_{i_2}$ for any fixed $k \in \mathbb{N}$. Because $\bar{N}_{i_1} \cap \bar{N}_{i_2}$ is compact, there exists a~subcover $\{ B(a_m, \tfrac{1}{k}) \}_{m \in F_k}$, where $F_k \subseteq \mathbb{N}$ is a~\emph{finite} set of indices.

We now claim that
\begin{align}
\label{Thm_p1}
& J^+_{(\bar{N}_{i_1}, \bar{N}_{i_2})} = \bigcap\limits_{k \in \mathbb{N}} \bigcup\limits_{m \in F_k} J^-_{(\bar{N}_{i_1})}\left(\bar{B}(a_m, \tfrac{1}{k})\right) \times J^+_{(\bar{N}_{i_2})}\left(\bar{B}(a_m, \tfrac{1}{k})\right)
\\
\nonumber
& = \left\{ (p,q) \in \bar{N}_{i_1} \times \bar{N}_{i_2} \ | \ \forall \, k \in \mathbb{N} \ \exists \, m \in F_k \ \exists \, p_m, q_m \in \bar{B}(a_m, \tfrac{1}{k}) \ \ p \preceq_{U_{i_1}} p_m \ \wedge \ q_m \preceq_{U_{i_2}} q \right\},
\end{align}
\noindent
which would already mean that $J^+_{(\bar{N}_{i_1}, \bar{N}_{i_2})}$ is a~closed subset of $\bar{N}_{i_1} \times \bar{N}_{i_2}$ (and hence also a~compact subset of $U_{i_1} \times U_{i_2}$), because finite unions of closed sets are closed and so are any intersections of closed sets.

Indeed, to prove the~inclusion `$\subseteq$', assume $(p,q) \in \bar{N}_{i_1} \times \bar{N}_{i_2}$ is such that there exists $r \in \bar{N}_{i_1} \cap \bar{N}_{i_2}$ satisfying $p \preceq_{U_{i_1}} r \preceq_{U_{i_2}} q$. For any $k \in \mathbb{N}$, since $\{ B(a_m, \tfrac{1}{k}) \}_{m \in F_k}$ covers $\bar{N}_{i_1} \cap \bar{N}_{i_2}$, it is possible to find $m \in F_k$ such that $r \in B(a_m, \tfrac{1}{k})$. One can thus simply take $p_m := r =: q_m$.

On the~other hand, to show the~inclusion `$\supseteq$', let us assume that $(p,q) \in \bar{N}_{i_1} \times \bar{N}_{i_2}$ is such that
\begin{align*}
\forall \, k \in \mathbb{N} \ \ \exists \, m \in F_k \ \ \exists \, p_m, q_m \in \bar{B}(a_m, \tfrac{1}{k}) \quad p \preceq_{U_{i_1}} p_m \ \textnormal{ and } \ q_m \preceq_{U_{i_2}} q.
\end{align*}
We can thus construct the~sequence $\{a_{m_k}\}_{k \in \mathbb{N}}$, which, being contained in the~compact set $\bar{N}_{i_1} \cap \bar{N}_{i_2}$, has a~subsequence $\{a_{m_{k_l}}\}_{l \in \mathbb{N}}$ convergent to some $a_{\infty} \in \bar{N}_{i_1} \cap \bar{N}_{i_2}$. Notice now that because $p_{m_k}, q_{m_k} \in \bar{B}(a_{m_k}, \tfrac{1}{k})$ for any $k \in \mathbb{N}$, therefore also
\begin{align*}
\lim\limits_{l \rightarrow +\infty} p_{m_{k_l}} = \lim\limits_{l \rightarrow +\infty} q_{m_{k_l}} = a_\infty.
\end{align*}
We now invoke the~fact that $J^+_{U_{i_1}}$ and $J^+_{U_{i_2}}$ are \emph{closed} subsets of $U_{i_1}^2$ and of $U_{i_2}^2$, respectively. It implies that
\begin{align*}
p \preceq_{U_{i_1}} a_\infty \preceq_{U_{i_2}} q,
\end{align*}
\noindent
which completes the~proof of (\ref{Thm_p1}) and of the~base case of the~induction.

We now move to the proof of the~inductive step, which essentially goes along the same lines as the proof of the~base case.

The~assumption is that $J^+_{(\bar{N}_{i_1}, \bar{N}_{i_2}, \ldots, \bar{N}_{i_n})}$ is a~compact subset of $U_{i_1} \times U_{i_n}$ for any $i_1, \ldots, i_n \in \mathbb{N}$.

The~induction hypothesis states that $J^+_{(\bar{N}_{i_1}, \bar{N}_{i_2}, \ldots, \bar{N}_{i_{n+1}})}$ is a~compact subset of $U_{i_1} \times U_{i_{n+1}}$ for any $i_1, \ldots, i_{n+1} \in \mathbb{N}$.

Let $\{a_m\}_{m \in \mathbb{N}}$ denote now a~dense subset of $N_{i_n}$, and hence also a~dense subset of $\bar{N}_{i_n}$. Similarly as before, for each $k \in \mathbb{N}$ consider the~family $\{ B(a_m, \tfrac{1}{k}) \}_{m \in \mathbb{N}}$ covering $\bar{N}_{i_n}$, and take its finite subcover $\{ B(a_m, \tfrac{1}{k}) \}_{m \in F_k}$.

We now claim that
\begin{align}
J^+_{(\bar{N}_{i_1}, \bar{N}_{i_2}, \ldots, \bar{N}_{i_{n+1}})}
& = \bigcap\limits_{k \in \mathbb{N}} \bigcup\limits_{m \in F_k} J^-_{(\bar{N}_{i_1}, \bar{N}_{i_2}, \ldots, \bar{N}_{i_n})}\left(\bar{B}(a_m, \tfrac{1}{k})\right) \times J^+_{(\bar{N}_{i_n}, \bar{N}_{i_{n+1}})}\left(\bar{B}(a_m, \tfrac{1}{k})\right) \label{Thm_p2}
\\
\nonumber
& = \Big\{ (p,q) \in \bar{N}_{i_1} \times \bar{N}_{i_{n+1}} \ | \ \forall \, k \in \mathbb{N} \ \exists \, m \in F_k \ \exists \, p_m, q_m \in \bar{B}(a_m, \tfrac{1}{k})
\\
\nonumber
& \hspace*{3.2cm}\ (p, p_m) \in J^+_{(\bar{N}_{i_1}, \bar{N}_{i_2}, \ldots, \bar{N}_{i_n})} \ \wedge \ (q_m, q) \in J^+_{(\bar{N}_{i_n}, \bar{N}_{i_{n+1}})} \Big\},
\end{align}
\noindent
which would already mean that $J^+_{(\bar{N}_{i_1}, \bar{N}_{i_2}, \ldots, \bar{N}_{i_{n+1}})}$ is a~closed subset of $\bar{N}_{i_1} \times \bar{N}_{i_{n+1}}$ (and hence also a~compact subset of $U_{i_1} \times U_{i_{n+1}}$), because we already know that $J^-_{(\bar{N}_{i_1}, \bar{N}_{i_2}, \ldots, \bar{N}_{i_n})}\left(\bar{B}(a_m, \tfrac{1}{k})\right)$ is closed in $\bar{N}_{i_1}$ (by the~induction assumption and definitions (\ref{fpsets5})) and that \linebreak $J^+_{(\bar{N}_{i_n}, \bar{N}_{i_{n+1}})}\left(\bar{B}(a_m, \tfrac{1}{k})\right)$ is closed in $\bar{N}_{i_{n+1}}$ (by the~base case and definitions (\ref{fpsets4})).

To show the~inclusion `$\subseteq$' in (\ref{Thm_p2}), assume $(p,q) \in \bar{N}_{i_1} \times \bar{N}_{i_{n+1}}$ is such that there exists $r \in \bar{N}_{i_n}$ satisfying $(p, r) \in J^+_{(\bar{N}_{i_1}, \bar{N}_{i_2}, \ldots, \bar{N}_{i_n})}$ and $(r, q) \in J^+_{(\bar{N}_{i_n}, \bar{N}_{i_{n+1}})}$. For any $k \in \mathbb{N}$, since $\{ B(a_m, \tfrac{1}{k}) \}_{m \in F_k}$ covers $\bar{N}_{i_n}$, it is possible to find $m \in F_k$ such that $r \in B(a_m, \tfrac{1}{k})$. One can thus simply take $p_m := r =: q_m$.

On the~other hand, to show the~inclusion `$\supseteq$', let us assume that $(p,q) \in \bar{N}_{i_1} \times \bar{N}_{i_{n+1}}$ are such that
\begin{align*}
\forall \, k \in \mathbb{N} \ \ \exists \, m \in F_k \ \ \exists \, p_m, q_m \in \bar{B}(a_m, \tfrac{1}{k}) \quad & (p, p_m) \in J^+_{(\bar{N}_{i_1}, \bar{N}_{i_2}, \ldots, \bar{N}_{i_n})}, \quad (q_m, q) \in J^+_{(\bar{N}_{i_n}, \bar{N}_{i_{n+1}})}.
\end{align*}
We can thus construct the~sequence $(a_{m_k})_{k \in \mathbb{N}}$, which, being contained in the~compact set $\bar{N}_{i_n}$, has a~subsequence $(a_{m_{k_l}})_{l \in \mathbb{N}}$ convergent to some $a_{\infty} \in \bar{N}_{i_n}$. Analogously as before, we argue that also the~sequences $(p_{m_k}), (q_{m_k})$ have subsequences converging to $a_\infty$.

We now invoke the~induction assumption and definitions (\ref{fpsets5}), which together imply that $J^+_{(\bar{N}_{i_1}, \bar{N}_{i_2}, \ldots, \bar{N}_{i_n})}$ is a~compact (and hence closed) subset of $U_{i_1} \times U_{i_n}$ and therefore $(p, a_\infty) \in J^+_{(\bar{N}_{i_1}, \bar{N}_{i_2}, \ldots, \bar{N}_{i_n})}$.

On the~other hand, invoking the~base case and definitions (\ref{fpsets4}), we also have that $J^+_{(\bar{N}_{i_n}, \bar{N}_{i_{n+1}})}$ is a~compact (and hence closed) subset of $U_{i_n} \times U_{i_{n+1}}$ and so $(a_\infty, q) \in J^+_{(\bar{N}_{i_n}, \bar{N}_{i_{n+1}})}$. This completes the~proof of (\ref{Thm_p2}) and of the~entire induction.

Altogether, we can thus write that
\begin{align}
\label{Thm_p3}
& \forall \, n \in \mathbb{N} \ \ \forall i_1, i_2, \ldots, i_n \in \mathbb{N}
\\
\nonumber
& \qquad J^+_{(\bar{N}_{i_1}, \bar{N}_{i_2}, \ldots, \bar{N}_{i_n})} \textnormal{ is a~compact subset of } U_{i_1} \times U_{i_n}, \textnormal{ and hence of } \M^2.
\end{align}

Bearing the~above in mind, the~$\sigma$-compactness of $J^+$ will be proven if we show that
\begin{align}
\label{Thm_p4}
J^+ = \bigcup\limits_{n = 1}^\infty \bigcup\limits_{i_1, i_2, \ldots, i_n \in \mathbb{N}} J^+_{(\bar{N}_{i_1}, \bar{N}_{i_2}, \ldots, \bar{N}_{i_n})}.
\end{align}

In order to show the~inclusion `$\subseteq$', take any $(p,q) \in J^+$ and let $\gamma: [0,1] \rightarrow \M$ be a~piecewise smooth future-directed causal curve from $p$ to $q$.

Consider the~inverse images $\gamma^{-1}(N_i)$, $i \in \mathbb{N}$. By continuity of $\gamma$, they are all open subsets of $[0,1]$, however they might be \emph{disconnected} (i.e. they need not be intervals). Nevertheless, every $\gamma^{-1}(N_i)$ is a~union of its connected components, which are all open\footnote{Locally compact spaces (and $[0,1]$ is such a~space) can be charaterised as the~spaces in which every connected component of every open set is itself open.} subintervals of $[0,1]$.

Let us thus consider the~family of all connected components of all $\gamma^{-1}(N_i)$'s, $i \in \mathbb{N}$. This family is a~cover of $[0,1]$ and, since the~latter is a~compact space, we can take its finite subcover ${\cal I} := \{I_1, I_2, \ldots, I_n\}$, where each of the~intervals $I_j$, ($j = 1,\ldots,n$) is a~connected component of some (possibly not unique) $\gamma^{-1}(N_{i_j})$. Therefore
\begin{align*}
\forall \, j = 1,\ldots,n \quad \gamma(I_j) \subseteq N_{i_j}
\end{align*}
\noindent
and, by the~continuity of $\gamma$,
\begin{align*}
\forall \, j = 1,\ldots,n \quad \gamma(\bar{I}_j) \subseteq \bar{N}_{i_j}.
\end{align*}
Without loss of generality, we can assume that $I_{j_1} \not\subseteq I_{j_2}$ for all $j_1 \neq j_2$. Bearing this in mind, we can rewrite ${\cal I}$ either as $\{[0,1]\}$ (the trivial cover) or, if $n > 1$, as
\begin{align*}
{\cal I} = \left\{ [0,b_1), (a_2,b_2), \ldots, (a_{n-1}, b_{n-1}), (a_n, 1] \right\},
\end{align*}
\noindent
where $0 < a_2 < a_3 < \ldots < a_n < 1$. Notice also that $b_j > a_{j+1}$ for $j=1,\ldots,n-1$, because otherwise such ${\cal I}$ would not be a~cover.

In the~first (trivial) case, $\gamma([0,1]) \subseteq N_{i_1} \subseteq \bar{N}_{i_1}$ for some $i_1 \in \mathbb{N}$ and hence $(p,q) \in J^+_{(\bar{N}_{i_1})}$.

In the~second case, observe that
\begin{align*}
& \gamma([0,a_2]) \subseteq \gamma([0,b_1)) \subseteq N_{i_1} \subseteq \bar{N}_{i_1},
\\
& \gamma([a_2,a_3]) \subseteq \gamma([a_2,b_2]) \subseteq \bar{N}_{i_2},
\\
& \ldots
\\
& \gamma([a_j,a_{j+1}]) \subseteq \gamma([a_j,b_j]) \subseteq \bar{N}_{i_j},
\\
& \ldots
\\
& \gamma([a_{n-1},a_n]) \subseteq \gamma([a_{n-1},b_{n-1}]) \subseteq \bar{N}_{i_{n-1}},
\\
& \gamma([a_n,1]) \subseteq \bar{N}_{i_n}\, ,
\end{align*}
\noindent
for some $i_1,\ldots,i_n \in \mathbb{N}$ and hence $(p,q) \in J^+_{(\bar{N}_{i_1},\ldots,\bar{N}_{i_n})}$.

In either case, we obtain that $(p,q) \in \bigcup\limits_{n = 1}^\infty \bigcup\limits_{i_1, i_2, \ldots, i_n \in \mathbb{N}} J^+_{(\bar{N}_{i_1}, \bar{N}_{i_2}, \ldots, \bar{N}_{i_n})}$.

In order to show the~other inclusion `$\supseteq$' in (\ref{Thm_p4}), notice simply that a~concatenation of finitely many piecewise smooth future-directed causal curves is itself a~piecewise smooth future-directed causal curve. Therefore, if $(p,q) \in J^+_{(\bar{N}_{i_1}, \bar{N}_{i_2}, \ldots, \bar{N}_{i_n})}$, then $(p,q) \in J^+$.
\end{Proof}

\begin{Cor}
Let $\M$ be a~spacetime. Then $E^+$ is a~$\sigma$-compact subset of $\M^2$.
\end{Cor}
\begin{Proof}\textbf{:}
On the~strength of (\ref{Thm_p4}), we have that
\begin{align*}
E^+ := J^+ \setminus I^+ = \bigcup\limits_{n = 1}^\infty \bigcup\limits_{i_1, i_2, \ldots, i_n \in \mathbb{N}} J^+_{(\bar{N}_{i_1}, \bar{N}_{i_2}, \ldots, \bar{N}_{i_n})} \setminus I^+
\end{align*}
\noindent
and since $I^+$ is an~open subset of $\M^2$, therefore $J^+_{(\bar{N}_{i_1}, \bar{N}_{i_2}, \ldots, \bar{N}_{i_n})} \setminus I^+$ is a~closed subset of $\bar{N}_{i_1} \times \bar{N}_{i_n}$ (for any $i_1, i_2, \ldots, i_n \in \mathbb{N}$), and hence a~compact subset of $\M^2$.
\end{Proof}

\begin{Cor}
Let $\M$ be a~spacetime and let $\X \subseteq \M$ be a~countable union of closed sets. Then $J^\pm(\X)$ and $E^\pm(\X)$ are $\sigma$-compact subsets of $\M$.
\end{Cor}
\begin{Proof}\textbf{:}
By assumption, $\X = \bigcup\limits_{m = 1}^\infty \X_m$, where for any $m \in \mathbb{N}$, $\X_m \subseteq \M$ is closed. Observe that, by (\ref{Thm_p4}),
\begin{align*}
J^+(\X) & := \pr_2\left( \left( \bigcup\limits_{m = 1}^\infty \X_m \times \M \right) \cap \bigcup\limits_{n = 1}^\infty \bigcup\limits_{i_1, i_2, \ldots, i_n \in \mathbb{N}} J^+_{(\bar{N}_{i_1}, \bar{N}_{i_2}, \ldots, \bar{N}_{i_n})} \right)
\\
& = \pr_2\left( \bigcup\limits_{m = 1}^\infty \bigcup\limits_{n = 1}^\infty \bigcup\limits_{i_1, i_2, \ldots, i_n \in \mathbb{N}} (\X_m \times \M) \cap J^+_{(\bar{N}_{i_1}, \bar{N}_{i_2}, \ldots, \bar{N}_{i_n})} \right)
\\
& = \bigcup\limits_{m = 1}^\infty \bigcup\limits_{n = 1}^\infty \bigcup\limits_{i_1, i_2, \ldots, i_n \in \mathbb{N}} \pr_2\left( (\X_m \times \M) \cap J^+_{(\bar{N}_{i_1}, \bar{N}_{i_2}, \ldots, \bar{N}_{i_n})} \right).
\end{align*}
\noindent
For any $m,n \in \mathbb{N}$ and any $i_1, i_2, \ldots, i_n \in \mathbb{N}$ the~set $(\X_m \times \M) \cap J^+_{(\bar{N}_{i_1}, \bar{N}_{i_2}, \ldots, \bar{N}_{i_n})}$ is closed in $\bar{N}_{i_1} \times \bar{N}_{i_n}$ and hence compact in $\M^2$. Since $\pr_2$ is a~continuous map, the~projection of a~compact set is itself compact and we obtain that $J^+(\X)$ is $\sigma$-compact.

The~proof for $J^-(\X)$ is completely analogous. Moreover, by the~previous corollary, replacing $J^\pm$ with $E^\pm$ in the~above proof yields the~desired result for the~horismotical futures and pasts.
\end{Proof}

The~final corollary shows that the~volume functions can be defined by means of causal futures instead of the~chronological ones.
\begin{Cor}
Let $\M$ be a~spacetime and $\eta \in \mathfrak{P}(\M)$ be an~admissible measure. Then the~volume functions $t^\pm$ associated to $\eta$ satisfy $t^\pm(p) = \mp \eta(J^\pm(p))$ for all $p \in \M$. Moreover, $\eta(E^\pm(p)) = 0$ for all $p \in \M$.
\end{Cor}
\begin{Proof}\textbf{:}
By the~previous corollary, $E^\pm(p)$ and $J^\pm(p)$ are Borel sets for any $p \in \M$ and so the~expressions $\eta(E^\pm(p))$ and $\eta(J^\pm(p))$ are well defined. Since it is true that
\begin{align*}
\forall \, p \in \M \quad I^-(p) \subseteq J^-(p) \subseteq \overline{J^-(p)} = \overline{I^-(p)} = I^-(p) \cup \partial I^-(p),
\end{align*}
\noindent
with $I^-(p) \cap \partial I^-(p) = \emptyset$, therefore
\begin{align*}
t^-(p) = \eta(I^-(p)) \leq \eta(J^-(p)) \leq \eta(\overline{J^-(p)}) = \eta(\overline{I^-(p)}) = \eta(I^-(p)) + \underbrace{\eta(\partial I^-(p))}_{= \, 0} = t^-(p),
\end{align*}
\noindent
where we have used the~second condition in the~definition of an~admissible measure. Therefore, $t^-(p) = \eta(J^-(p))$. The~proof for $t^+$ is analogous.

Moreover, since $I^\pm(p) \subseteq J^\pm(p)$ for any $p \in \M$, therefore $\eta(E^\pm(p)) = \eta(J^\pm(p) \setminus I^\pm(p)) = \eta(J^\pm(p)) - \eta(I^\pm(p)) = 0$.
\end{Proof}

\section{Causality for probability measures}\label{sec:main}

The~aim of this section is to extend the~causal precedence relation $\preceq$ onto the~space of measures $\mathfrak{P}(\M)$ on a~given spacetime $\M$. We begin by invoking certain characterisation of causality between events.

Let ${\cal C}(\M)$ denote the~set of smooth bounded causal functions on the~spacetime $\M$.
\begin{Thm}
\label{causalfunctions}
Let $\M$ be a~globally hyperbolic spacetime. For any $p,q \in \M$ the~following conditions are equivalent
\begin{enumerate}[1{$^\diamond$}]
\item $\forall \, f \in {\cal C}(\M) \quad f(p) \leq f(q)$,
\item $p \preceq q$.
\end{enumerate}
\end{Thm}
The proof, based on a result by Besnard \cite{Bes09}, can be found in \cite[Proposition 10]{CQG2013} (see also \cite{MinguzziUtilities}). Actually, as we shall see later, the~above characterisation is valid also in causally simple spacetimes (cf. Corollary \ref{couplingsCor2}).

As an~important side note, observe that Theorem \ref{causalfunctions} exactly mirrors the~definition of a~causal function. Indeed, the~latter can be written symbolically as
\begin{align*}
f \textnormal{ a~causal function iff } \ \forall (p,q) \in J^+ \ \, f(p) \leq f(q),
\end{align*}
\noindent
whereas Theorem \ref{causalfunctions} in fact says that
\begin{align*}
(p,q) \in J^+ \textnormal{ iff } \ \forall \, f \textnormal{ a~causal function } \ \, f(p) \leq f(q).
\end{align*}
\noindent
Therefore, instead of using $\preceq$ to define what a~causal function is, one can come up with an~abstract, suitably structurised set ${\cal C}$ of `smooth bounded causal functions' and \emph{define} $\preceq$ through ${\cal C}$ using the~analogue of Theorem \ref{causalfunctions}. This was done by Eckstein and Franco in \cite{CQG2013} in very general context of noncommutative geometry.

Condition $1^\diamond$ provides a~`dual' definition of the~causal precedence, which actually suggests how $\preceq$ could be extended onto $\mathfrak{P}(\M).$
\begin{Def}
\label{causality_def1}
Let $\M$ be a~globally hyperbolic spacetime. For any $\mu,\nu \in \mathfrak{P}(\M)$ we say that $\mu$ \emph{causally precedes} $\nu$ (symbolically $\mu \preceq \nu$) iff
\begin{align*}
\forall \, f \in {\cal C}(\M) \quad \int\limits_{\M} f \, d\mu \leq \int\limits_{\M} f \, d\nu.
\end{align*}
\end{Def}
In \cite{CQG2013} it is proven (in a~much more general context) that the~above defined relation is in fact a~partial order. This definition, however, has two shortcomings. Firstly, it is well motivated only on globally hyperbolic spacetimes. Secondly, the intuitive notion of causality for spread objects, as phrased in the introduction, is not directly visible in Definition \ref{causality_def1}.

\subsection{Characterisations of the~causal relation}\label{subsec:characterisation}

In the~following, we provide various conditions which are equivalent to the~above definition of a causal relation between measures. Moreover, in some of the implications the assumption on global hyperbolicity of $\M$ can be relaxed.

The~first result states that if ${\cal C}(\M)$ is sufficiently rich, one can abandon the~smoothness requirement.
\begin{Thm}
\label{main_result1}
Let $\M$ be a~stably causal spacetime. For any $\mu, \nu \in \mathfrak{P}(\M)$ the~following conditions are equivalent:
\begin{enumerate}[1{$^\bullet$}]
\item For all $f \in {\cal C}(\M)$
    \begin{align}
    \label{causal1}
    \int\limits_{\M} f \, d\mu \leq \int\limits_{\M} f \, d\nu.
    \end{align}
\item For all causal $f \in C_b(\M)$
    \begin{align}
    \label{causal2}
    \int\limits_{\M} f \, d\mu \leq \int\limits_{\M} f \, d\nu.
    \end{align}
\end{enumerate}
\end{Thm}
\begin{Proof}\textbf{:}
($1^\bullet \Rightarrow 2^\bullet$) Relying on \cite[Corollary 5.4 and the~subsequent comments]{ChruscielDiff} we use the~fact that in stably causal spacetimes any time function can be uniformly approximated by a~\emph{smooth} time (or even temporal) function.

Using the~stable causality, fix a~temporal function ${\cal T}: \M \rightarrow \mathbb{R}$. For any $\varepsilon > 0$, the function $f + \varepsilon \arctan {\cal T}$ is a~time function which clearly approximates $f$ uniformly. By the~above mentioned corollary, this function in turn can be approximated by a~smooth time function $f_\varepsilon$ such that
\begin{align}
\label{uniform_approx}
\forall \, p \in \M \quad \left|f(p) + \varepsilon \arctan {\cal T}(p) - f_\varepsilon(p) \right| < \varepsilon.
\end{align}
Clearly $f_\varepsilon \in {\cal C}(\M)$, therefore by $1^\bullet$
\begin{align*}
\int\limits_{\M} f_\varepsilon \, d\mu \leq \int\limits_{\M} f_\varepsilon \, d\nu.
\end{align*}
To obtain $2^\bullet$ it now remains to observe that for any measure $\eta \in \mathfrak{P}(\M)$ it is true that $\lim\limits_{\varepsilon \rightarrow 0^+} \int\limits_{\M} f_\varepsilon \, d\eta = \int\limits_{\M} f \, d\eta$.

Indeed, for any $\eta \in \mathfrak{P}(\M)$ and $\varepsilon > 0$ one has
\begin{align*}
\left| \int\limits_{\M} f \, d\eta - \int\limits_{\M} f_\varepsilon \, d\eta \right| & \leq \int\limits_{\M} |f - f_\varepsilon| d\eta \leq \int\limits_{\M} \left|f + \varepsilon \arctan {\cal T} - f_\varepsilon \right| d\eta + \varepsilon \int\limits_{\M} | \arctan {\cal T} | d\eta
\\
& \leq \varepsilon \left( 1 + \tfrac{\pi}{2} \right),
\end{align*}
\noindent
where we have used (\ref{uniform_approx}).

($2^\bullet \Rightarrow 1^\bullet$) Trivial.
\end{Proof}

The~next result characterises the relation $\preceq$ between measures in terms of open future sets.

\begin{Thm}
\label{main_result2}
Let $\M$ be a~causally continuous spacetime. For any $\mu, \nu \in \mathfrak{P}(\M)$ conditions $1^\bullet$ and $2^\bullet$ are equivalent to the~following condition
\begin{enumerate}[1{$^\bullet$}]
\setcounter{enumi}{2}
\item For every open future set ${\cal F} \subseteq \M$
\begin{align}
    \label{causal3}
    \mu({\cal F}) \leq \nu({\cal F}).
\end{align}
\end{enumerate}
\end{Thm}
\begin{Proof}\textbf{:}
($2^\bullet \Rightarrow 3^\bullet$) Fix an~open future set ${\cal F} \subseteq \M$ and let $\eta$ be an~admissible measure on $\M$. For any $\lambda \in (0,1]$ construct a~new admissible measure $\eta_\lambda := \lambda \eta + (1-\lambda)\eta(\, \cdot \cap {\cal F})$ and consider the~associated past volume function $t^-_\lambda$ defined via
\begin{align*}
\forall \, p \in \M \quad t^-_\lambda(p) & := \eta_\lambda(I^-(p)) = \lambda \eta(I^-(p)) + (1-\lambda)\eta(I^-(p) \cap {\cal F})
\\
& = \eta(I^-(p) \cap {\cal F}) + \lambda \eta(I^-(p) \setminus {\cal F}).
\end{align*}
\noindent
Because $\M$ is causally continuous, $t^-_\lambda$ is a~time function for any $\lambda \in (0,1]$.

Now, for every $n \in \mathbb{N}$ define an~increasing function $\varphi_n \in C^\infty_b(\mathbb{R})$ by
\begin{align*}
\forall \, x \in \mathbb{R} \quad \varphi_n(x) := \tfrac{1}{2} + \tfrac{1}{\pi} \arctan \left( n^2 x - n \right).
\end{align*}
The~sequence of functions $(\varphi_n)$ is pointwise convergent to the~indicator function of $\mathbb{R}_{>0}$. Moreover, also $(\varphi_n \circ t^-_\lambda)$ is a~bounded time function for every $n \in \mathbb{N}$ and $\lambda \in (0,1]$. By $2^\bullet$, this means that
\begin{align*}
\int\limits_{\M} \varphi_n\left( \eta(I^-(p) \cap {\cal F}) + \lambda \eta(I^-(p) \setminus {\cal F}) \right) \, d\mu(p) \leq \int\limits_{\M} \varphi_n\left( \eta(I^-(p) \cap {\cal F}) + \lambda \eta(I^-(p) \setminus {\cal F}) \right) \, d\nu(p).
\end{align*}

Since the~functions $\varphi_n$ are bounded and continuous, we can invoke Lebesgue's dominated convergence theorem and first take $\lambda \rightarrow 0^+$, obtaining
\begin{align*}
\int\limits_{\M} \varphi_n\left( \eta(I^-(p) \cap {\cal F}) \right) \, d\mu(p) \leq \int\limits_{\M} \varphi_n\left( \eta(I^-(p) \cap {\cal F}) \right) \, d\nu(p)
\end{align*}
\noindent
and then take $n \rightarrow +\infty$, which yields
\begin{align*}
\int\limits_{\M} \jdn_{\mathbb{R}_{>0}}\left( \eta(I^-(p) \cap {\cal F}) \right) \, d\mu(p) \leq \int\limits_{\M} \jdn_{\mathbb{R}_{>0}}\left( \eta(I^-(p) \cap {\cal F}) \right) \, d\nu(p) \, .
\end{align*}
It is now crucial to notice that the~function $p \mapsto \eta({\cal F} \cap I^-(p))$ is \emph{positive} on ${\cal F}$ and \emph{zero} on $\M \setminus {\cal F}$. These observations follow from the~definition of an~admissible measure and of ${\cal F}$, and together with the~above inequality of integrals they imply that
\begin{align*}
    \mu({\cal F}) \leq \nu({\cal F}).
\end{align*}

($3^\bullet \Rightarrow 2^\bullet$) Let $f \in C_b(\M)$ be causal and let ${\cal T}$ be a~temporal function\footnote{Such a function exists because causal continuity implies stable causality. In fact, in the~proof of ($3^\bullet \Rightarrow 2^\bullet$) we only need $\M$ be stably causal.} on $\M$. For any $\varepsilon > 0$ define a~bounded time function $f_\varepsilon := f + \varepsilon \arctan {\cal T}$.


Denote $m := \inf\limits_{p \in \M} f_\varepsilon(p)$ and $M := \sup\limits_{p \in \M} f_\varepsilon(p)$. For any fixed $n \in \mathbb{N}$ define the~sets
\begin{align*}
{\cal F}_k^{(n)} := f_\varepsilon^{-1} \left( \left( m + k \tfrac{M-m}{n}, +\infty \right) \right), \qquad k=1,2,\ldots,n-1.
\end{align*}
Because $f_\varepsilon$ is continuous and causal, all ${\cal F}_k^{(n)}$'s are open future sets (cf. Proposition \ref{prelimLem3}).

For any fixed $n \in \mathbb{N}$ let us consider the~following simple function
\begin{align*}
s_n := m + \sum\limits_{k=1}^{n-1} \tfrac{M-m}{n} \jdn_{{\cal F}_k^{(n)}}.
\end{align*}

By $3^\bullet$, we obtain the~following inequality of integrals
\begin{align}
\label{3then2a}
\int\limits_{\M} s_n \, d\mu = m + \sum\limits_{k=1}^{n-1} \tfrac{M-m}{n} \mu\left( {\cal F}_k^{(n)} \right) \leq m + \sum\limits_{k=1}^{n-1} \tfrac{M-m}{n} \nu\left( {\cal F}_k^{(n)} \right) = \int\limits_{\M} s_n \, d\nu.
\end{align}
It is not difficult to realise that
\begin{align*}
\forall \, p \in \M \quad \left[ \forall n \in \mathbb{N} \ \ s_n(p) < f_\varepsilon(p) \right] \ \textrm{and} \ \lim\limits_{n \rightarrow +\infty} s_n(p) = f_\varepsilon(p).
\end{align*}
More concretely, one can show that
\begin{align}
\label{3then2b}
\forall \, p \in \M \quad f_\varepsilon(p) - s_n(p) \in (0,\tfrac{M-m}{n}].
\end{align}
Indeed, the~very definition of ${\cal F}_k^{(n)}$'s implies that ${\cal F}_1^{(n)} \supset {\cal F}_2^{(n)} \supset \ldots \supset {\cal F}_{n-1}^{(n)}$, therefore if $p \in {\cal F}_k^{(n)}$ for some $k \in \{1,\ldots,n-1\}$, then $p \in {\cal F}_j^{(n)}$ for all $j \in \{1,\ldots,k\}$. This implies that
\begin{align*}
s_n(p) & = m + \sum\limits_{k=1}^{n-1} \tfrac{M-m}{n} \jdn_{{\cal F}_k^{(n)}}(p) = m + \tfrac{M-m}{n} \max \left\{ k \ | \ p \in {\cal F}_k^{(n)} \right\}
\\
& = m + \tfrac{M-m}{n} \max \left\{ k \ | \ m + k \tfrac{M-m}{n} < f_\varepsilon(p) \right\}
\\
& = m + \tfrac{M-m}{n} \max \left\{ k \ | \ k < \tfrac{n}{M-m} (f_\varepsilon(p) - m) \right\}
\\
& = m + \tfrac{M-m}{n} \left\lceil \tfrac{n}{M-m} (f_\varepsilon(p) - m) - 1 \right\rceil,
\end{align*}
\noindent
where $\lceil \cdot \rceil$ denotes the~ceiling function. Using the~fact that $x - \lceil x - 1 \rceil \in (0,1]$ for any $x \in \mathbb{R}$, we obtain that
\begin{align*}
f_\varepsilon(p) - s_n(p) = \tfrac{M-m}{n} \Big( \tfrac{n}{M-m} (f_\varepsilon(p) - m) - \left\lceil \tfrac{n}{M-m} (f_\varepsilon(p) - m) - 1 \right\rceil \Big) \in \left(0,\tfrac{M-m}{n}\right],
\end{align*}
\noindent
which proves (\ref{3then2b}).

Invoking now Lebesgue's dominated convergence theorem and passing with $n \rightarrow +\infty$ in (\ref{3then2a}) we obtain
\begin{align*}
\int\limits_{\M} f_\varepsilon \, d\mu \leq \int\limits_{\M} f_\varepsilon \, d\nu.
\end{align*}

Invoking Lebesgue's theorem again, we pass with $\varepsilon \rightarrow 0^+$ and obtain $2^\bullet$.
\end{Proof}

The~third and the~most important result concerns causally simple spacetimes. We show that condition $3^\bullet$ extends to different kinds of future sets. Moreover, we introduce a~condition that uses the~existential quantifier.
\begin{Thm}
\label{main_result3}
Let $\M$ be a~causally simple spacetime. For any $\mu, \nu \in \mathfrak{P}(\M)$ conditions $1^\bullet$, $2^\bullet$ and $3^\bullet$ are equivalent to all the~following conditions
\begin{enumerate}[1{$^\bullet$}]
\setcounter{enumi}{3}
\item For every compact $\K \subseteq \M$
    \begin{align}
    \label{causal4}
    \mu(J^+(\K)) \leq \nu(J^+(\K)).
    \end{align}
\item For every Borel future set ${\cal F} \subseteq \M$
\begin{align}
    \label{causal5}
    \mu({\cal F}) \leq \nu({\cal F}).
\end{align}
\item For all $\varphi, \psi \in C_b(\M)$
    \begin{align}
    \label{causal6}
    \big[ \forall p,q \in \M \quad p \preceq q \ \Rightarrow \ \varphi(p) \leq \psi(q) \big] \ \quad \Rightarrow \quad \ \int\limits_{\M} \varphi \, d\mu \leq \int\limits_{\M} \psi \, d\nu.
    \end{align}
\item There exists $\omega \in \mathfrak{P}(\M^2)$ such that
\begin{enumerate}[i)]
\item $(\pr_1)_\ast \omega = \mu$ and $(\pr_2)_\ast \omega = \nu$;
\item $\omega(J^+) = 1$.
\end{enumerate}
\end{enumerate}
\end{Thm}
\begin{Proof}\textbf{:}

($3^\bullet \Rightarrow 4^\bullet$) Let $\K$ be a~compact subset of $\M$. Fix $n \in \mathbb{N}$ and cover $\K$ with open balls of radius $\tfrac{1}{n}$, concretely
\begin{align*}
\K \subseteq \bigcup\limits_{x \in \K} B\left(x, \tfrac{1}{n}\right).
\end{align*}

\begin{figure}[h]
\begin{center}
\psfrag{K}{\scriptsize $\K$}
\psfrag{J}{\scriptsize $\bigcup\limits_{x \in \K} J^+ \big( B\left(x, \tfrac{1}{n}\right) \big)$}
\resizebox{!}{150pt}{\includegraphics[scale=0.6]{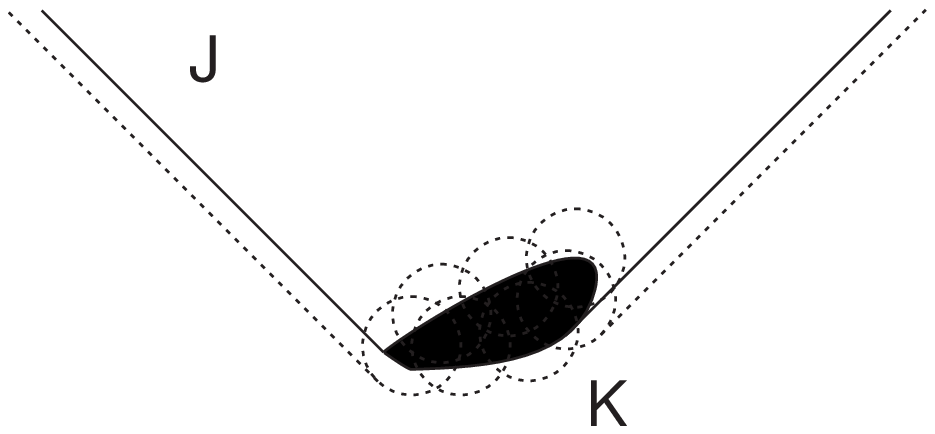}}
\caption{$\left\{J^+ \big( B\left(x, \tfrac{1}{n}\right) \big)\right\}_{x \in \K}$ covers $J^+(\K)$.}
\end{center}
\end{figure}

Hence
\begin{align}
\label{3to4a}
\forall \, n \in \mathbb{N} \quad J^+(\K) \subseteq J^+\left(\bigcup\limits_{x \in \K} B\left(x, \tfrac{1}{n}\right)\right) = \bigcup\limits_{x \in \K} J^+ \left(B\left(x, \tfrac{1}{n}\right)\right).
\end{align}
We claim that
\begin{align}
\label{3to4b}
J^+(\K) = \bigcap\limits_{n = 1}^{\infty} \bigcup\limits_{x \in \K} J^+ \left(B\left(x, \tfrac{1}{n}\right)\right).
\end{align}
By (\ref{3to4a}), it suffices to prove the~inclusion `$\supseteq$'.

Suppose then that $q \in \bigcap\limits_{n = 1}^{\infty} \bigcup\limits_{x \in \K} J^+ \left(B\left(x, \tfrac{1}{n}\right)\right)$, which means that
\begin{align*}
\forall \, n \in \mathbb{N} \ \ \exists \, x_n \in \K \ \ \exists \, p_n \in B(x_n, \tfrac{1}{n}) \ \quad p_n \preceq q.
\end{align*}
Since $\K$ is compact, the~sequence $(x_n)$ has a~convergent subsequence $(x_{n_k})$, $\lim\limits_{k \rightarrow +\infty} x_{n_k} = x_\infty \in \K$. Notice that also the~subsequence $(p_{n_k})$ converges to $x_\infty$. But because $J^+$ is a~closed set in the~case of a~causally simple spacetime, the~fact that for every $k \in \mathbb{N}$ $p_{n_k} \preceq q$ implies that $x_\infty \preceq q$ and therefore $q \in J^+(\K)$.

By $3^\bullet$ we know that
\begin{multline}
\label{3to4c}
\mu \left( \bigcup\limits_{x \in \K} J^+ \left(B\left(x, \tfrac{1}{n}\right)\right) \right) = \mu \left( \bigcup\limits_{x \in \K} I^+ \left(B\left(x, \tfrac{1}{n}\right)\right) \right)
\\
\leq \nu \left( \bigcup\limits_{x \in \K} I^+ \left(B\left(x, \tfrac{1}{n}\right)\right) \right) = \nu \left( \bigcup\limits_{x \in \K} J^+ \left(B\left(x, \tfrac{1}{n}\right)\right) \right)
\end{multline}

Since for all $n \in \mathbb{N}$ it is true that $J^+ \left(B\left(x, \tfrac{1}{n}\right)\right) \supseteq J^+ \left(B\left(x, \tfrac{1}{n+1}\right)\right)$, therefore, by (\ref{meas1}),
\begin{align*}
\mu\left( J^+(\K)\right) & = \mu \left( \bigcap\limits_{n = 1}^{\infty} \bigcup\limits_{x \in \K} J^+ \left(B\left(x, \tfrac{1}{n}\right)\right) \right) = \lim\limits_{n \rightarrow +\infty} \mu \left( \bigcup\limits_{x \in \K} J^+ \left(B\left(x, \tfrac{1}{n}\right)\right) \right)
\\
& \leq \lim\limits_{n \rightarrow +\infty} \nu \left( \bigcup\limits_{x \in \K} J^+ \left(B\left(x, \tfrac{1}{n}\right)\right) \right) = \nu \left( \bigcap\limits_{n = 1}^{\infty} \bigcup\limits_{x \in \K} J^+ \left(B\left(x, \tfrac{1}{n}\right)\right) \right) = \nu\left( J^+(\K)\right),
\end{align*}
\noindent
where we have also used (\ref{3to4b}) and (\ref{3to4c}), thus proving $4^\bullet$.
\\

($4^\bullet \Rightarrow 5^\bullet$) Let ${\cal F} \subseteq \M$ be any Borel future set. For any $\K \subseteq {\cal F}$ it is then true that $J^+(\K) \subseteq {\cal F}$. Therefore
\begin{align*}
\mu(\K) \leq \mu(J^+(\K)) \leq \mu({\cal F}).
\end{align*}
\noindent
In the~above chain of inequalities let us take the~supremum over all compact $\K \subseteq {\cal F}$. Using the~tightness of $\mu$ (see \eqref{meas4}), we have
\begin{align*}
\mu({\cal F}) & = \sup \left\{ \mu(\K) \ | \ \K \subseteq {\cal F}, \K \textnormal{ compact} \right\} \leq \sup \left\{ \mu(J^+(\K)) \ | \ \K \subseteq {\cal F}, \K \textnormal{ compact} \right\} \leq \mu({\cal F}),
\end{align*}
\noindent
and so
\begin{align*}
\mu({\cal F}) & = \sup \left\{ \mu(J^+(\K)) \ | \ \K \subseteq {\cal F}, \K \textnormal{ compact} \right\}
\end{align*}
\noindent
and similarly for the measure $\nu$. As we can see, in order to obtain $5^\bullet$ from $4^\bullet$ it is enough to take the~supremum over all compact $\K \subseteq {\cal F}$.
\\

($5^\bullet \Rightarrow 3^\bullet$) Trivial --- open sets are Borel.
\\

($2^\bullet \Rightarrow 6^\bullet$) In the~first step of the~proof we will show that $6^\bullet$ holds for all \emph{nonnegative} $\varphi, \psi \in C_b(\M)$ \emph{with} $\varphi$ \emph{compactly supported}. Namely, for such functions we will show that the~condition
\begin{align}
\label{Kantorovich1}
\forall p,q \in \M \quad p \preceq q \ \Rightarrow \ \varphi(p) \leq \psi(q)
\end{align}
\noindent
implies the~inequality of integrals
\begin{align}
\label{Kantorovich2}
\int\limits_{\M} \varphi \, d\mu \leq \int\limits_{\M} \psi \, d\nu.
\end{align}
Then, in the~second step, we will demonstrate that the~assumptions of nonnegativity of $\varphi, \psi$ and of the~compactness of $\supp \, \varphi$ can in fact be abandoned.

Define a~function $\hat{\varphi}: \M \rightarrow \mathbb{R}$ via $\hat{\varphi}(p) := \max\limits_{x \preceq p} \varphi(x)$. Function $\hat{\varphi}$ is well-defined, because for every $p \in \M$ the~function $\varphi$, being continuous, attains its maximum over the~compact\footnote{We are using the~fact that in causally simple spacetimes $J^\pm(p)$ are closed sets for all $p \in \M$.} set $J^-(p) \cap \supp \, \varphi$. Moreover, $\hat{\varphi}$ satisfies
\begin{align}
\label{Kantorovich3}
\forall p_1,p_2,q \in \M \quad p_1 \preceq p_2 \preceq q \ \Rightarrow \ \varphi(p_1) \leq \hat{\varphi}(p_2) \leq \psi(q)
\end{align}
Indeed, first inequality follows directly from the~very definition of $\hat{\varphi}$. In order to obtain the~second inequality, notice first that by (\ref{Kantorovich1}) we have $\varphi(p_2) \leq \psi(q)$. By transitivity of the~relation $\preceq$, this inequality holds also if we replace $p_2$ with any $x \preceq p_2$. Hence
\begin{align*}
\hat{\varphi}(p_2) = \max\limits_{x \preceq p_2} \varphi(x) \leq \psi(q)
\end{align*}
\noindent
and (\ref{Kantorovich3}) is proven.

Function $\hat{\varphi}$ is obviously nonnegative, bounded and, by transitivity of $\preceq$, it is causal. We claim that it is also continuous.

Indeed, let us show that for any $\alpha, \beta \in \mathbb{R}$ ($\alpha < \beta$) the~preimage $\hat{\varphi}^{-1}((\alpha,\beta))$ is open.

Notice first that if $\beta \leq 0$ then, by nonnegativity of $\hat{\varphi}$, the~preimage $\hat{\varphi}^{-1}((\alpha,\beta))$ is empty and hence open. Therefore, we can assume from now on that $\beta > 0$.

Observe that $\hat{\varphi}^{-1}((\alpha,+\infty)) = I^+\left( \varphi^{-1}((\alpha,+\infty)) \right)$. This is proven by the~following chain of equivalences
\begin{align*}
p \in \hat{\varphi}^{-1}((\alpha,+\infty)) & \quad \Leftrightarrow \quad \hat{\varphi}(p) > \alpha \quad \Leftrightarrow \quad \max\limits_{x \preceq p} \varphi(x) > \alpha \quad \Leftrightarrow \quad \exists \, x \preceq p \quad \varphi(x) > \alpha
\\
& \quad \Leftrightarrow \quad \exists \, x \in \varphi^{-1}((\alpha,+\infty)) \quad x \preceq p \quad \Leftrightarrow \quad p \in J^+\left( \varphi^{-1}((\alpha,+\infty)) \right)
\end{align*}
\noindent
and by the observation that, because $\varphi$ is continuous, $\varphi^{-1}((\alpha,+\infty))$ is open and hence $J^+\left( \varphi^{-1}((\alpha,+\infty)) \right) = I^+\left( \varphi^{-1}((\alpha,+\infty)) \right)$.

Similarly, observe that $\hat{\varphi}^{-1}([\beta,+\infty)) = J^+\left( \varphi^{-1}([\beta,+\infty)) \right)$. This is proven by a~chain of equivalences analogous to the~one above. Notice that because $\varphi$ is continuous, the~preimage $\varphi^{-1}([\beta,+\infty))$ is closed. Moreover, since $\varphi$ is nonnegative and $\beta > 0$, therefore $\varphi^{-1}([\beta,+\infty))$ is contained in the~support of $\varphi$. But the~latter is compact, and so the~preimage $\varphi^{-1}([\beta,+\infty))$, being a~closed subset of a~compact set, is itself compact. By the~causal simplicity of $\M$, the~set $J^+\left( \varphi^{-1}([\beta,+\infty)) \right)$ is closed.

Finally, notice that
\begin{align*}
\hat{\varphi}^{-1}((\alpha,\beta)) = \hat{\varphi}^{-1}((\alpha,+\infty)) \setminus \hat{\varphi}^{-1}([\beta,+\infty)) = I^+\left( \varphi^{-1}((\alpha,+\infty)) \right) \ \setminus \ J^+\left( \varphi^{-1}([\beta,+\infty)) \right)
\end{align*}
\noindent
which proves that $\hat{\varphi}^{-1}((\alpha,\beta))$ is an~open set.

We have thus shown that $\hat{\varphi} \in C_b(\M)$. By $2^\bullet$ we have that
\begin{align}
\label{aim}
\int\limits_{\M} \hat{\varphi} \, d \mu \leq \int\limits_{\M} \hat{\varphi} \, d \nu.
\end{align}
But from (\ref{aim}) we readily obtain (\ref{Kantorovich2}), because
\begin{align*}
\int\limits_{\M} \varphi \, d\mu \leq \int\limits_{\M} \hat{\varphi} \, d\mu \leq \int\limits_{\M} \hat{\varphi} \, d\nu \leq \int\limits_{\M} \psi \, d\nu,
\end{align*}
\noindent
where the~first and the~last inequalities follow from (\ref{Kantorovich3}) and the~middle one is exactly (\ref{aim}).

Thus, we have already proven $6^\bullet$ under the~assumption that $\varphi$ is compactly supported and both $\varphi$ and $\psi$ are nonnegative. Let us now take any $\varphi, \psi \in C_b(\M)$ satisfying (\ref{Kantorovich1}).

Define $m := \min \{ \inf \varphi, \inf \psi \}$ and introduce $\varphi_m, \psi_m \in C_b(\M)$ as $\varphi_m := \varphi - m$ and $\psi_m := \psi - m$. Of course $\varphi_m, \psi_m \geq 0$.

Let $(K_n)_{n \in \mathbb{N}}$ be an~exhaustion of $\M$ by compact sets. Using Urysohn's lemma for LCH spaces (Theorem \ref{UrysohnLCH}), we construct a~sequence $(\theta_n)_{n \in \mathbb{N}} \subseteq C_c(\M)$ of functions such that, for any $n \in \mathbb{N}$, $\theta_n|_{K_n} \equiv 1$ and $0 \leq \theta_n \leq 1$.

Notice that (for every $n \in \mathbb{N}$) the~function $\theta_n \varphi_m$ is compactly supported and, together with $\psi_m$, they are nonnegative and satisfy (\ref{Kantorovich1}), because for all $p,q \in \M$ such that $p \preceq q$ one has
\begin{align*}
\theta_n(p) \varphi_m(p) \leq \varphi_m(p) = \varphi(p) - m \leq \psi(q) - m = \psi_m(q).
\end{align*}
On the~strength of the~previous part of the~proof, it is then true that
\begin{align}
\label{Kantorovich7}
\int\limits_{\M} \theta_n \varphi_m \, d\mu \leq \int\limits_{\M} \psi_m \, d\nu.
\end{align}
By the~very definition, $\theta_n \leq 1$ for every $n$ and, since $(K_n)_{n \in \mathbb{N}}$ exhausts $\M$, we have that $\theta_n \rightarrow 1$ pointwise. By Lebesgue's dominated convergence theorem we can pass with $n \rightarrow +\infty$ in (\ref{Kantorovich7}) obtaining
\begin{align*}
\int\limits_{\M} \varphi_m \, d\mu \leq \int\limits_{\M} \psi_m \, d\nu.
\end{align*}
This, in turn, yields
\begin{align*}
\int\limits_{\M} (\varphi(p) - m) \, d\mu(p) \leq \int\limits_{\M} (\psi(q) - m) \, d\nu(q),
\end{align*}
\noindent
which, by the~fact that $\mu, \nu$ are probability measures, simplifies to
\begin{align*}
\int\limits_{\M} \varphi \, d\mu \leq \int\limits_{\M} \psi \, d\nu
\end{align*}
\noindent
and the~proof of $6^\bullet$ is complete.
\\

($6^\bullet \Rightarrow 7^\bullet$) We will use one of the~classical results in the~optimal transport theory, concerning what is known as the~Kantorovich duality. Concretely, we need the~following result adapted from \cite[Theorem 1.3]{CV03}.
\begin{Thm} (Kantorovich duality)
\label{Kantorovich_duality}
Let $({\cal X}_1, \mu_1)$ and $({\cal X}_2, \mu_2)$ be two Polish probability spaces and let $c: {\cal X}_1 \times {\cal X}_2 \rightarrow \mathbb{R}_{\geq 0} \cup \{ +\infty \}$ be a~lower semi-continuous function. Then
\begin{align}
\label{duality}
\min\limits_{\pi \in \Pi(\mu_1,\mu_2)} \int\limits_{{\cal X}_1 \times {\cal X}_2} c \, d\pi = \sup\limits_{(\varphi,\psi) \in \Psi(\mu_1,\mu_2)} \left( \int\limits_{{\cal X}_1} \varphi\, d\mu_1 - \int\limits_{{\cal X}_2} \psi\, d\mu_2 \right),
\end{align}
\noindent
where
\begin{itemize}
\item $\Pi(\mu_1,\mu_2) := \left\{ \pi \in \mathfrak{P}({\cal X}_1 \times {\cal X}_2) \ | \ (\pr_i)_\ast \pi = \mu_i, \ i=1,2 \right\}$,
\item $\Psi(\mu_1,\mu_2) := \{ (\varphi,\psi) \in C_b({\cal X}_1) \times C_b({\cal X}_2) \ | \ \forall x \in {\cal X}_1 \ \forall y \in {\cal X}_2 \quad \varphi(x) - \psi(y) \leq c(x,y) \}$.
\end{itemize}
\end{Thm}

Let us apply the~above theorem to the~setting in which $({\cal X}_1, \mu_1) := (\M, \mu)$, $({\cal X}_2, \mu_2) := (\M, \nu)$ and $c: \M^2 \rightarrow \mathbb{R}_{\geq 0} \cup \{ +\infty \}$ is defined as
\begin{align*}
c(p,q) = \left\{ \begin{array}{ll}
                 0 & \textrm{if } p \preceq q
                 \\
                 +\infty & \textrm{if } p \not\preceq q
                 \end{array} \right. \cdot
\end{align*}

The~assumptions of Theorem \ref{Kantorovich_duality} are met. $\M$ is a~Polish space (cf. Section \ref{subsec:topology}), whereas the~function $c$ is lower semi-continuous, because the~causal simplicity of $\M$ implies that $J^+$ is a~\emph{closed} subset of $\M^2$.

Notice that in the~above setting
\begin{align*}
\Psi(\mu,\nu) = \{ (\varphi,\psi) \in C_b(\M) \times C_b(\M) \ | \ \forall p,q \in \M \quad p \preceq q \ \Rightarrow \ \varphi(p) - \psi(q) \leq 0 \}.
\end{align*}
In other words, $\Psi(\mu,\nu)$ is the~set of exactly these pairs of functions which satisfy the~assumptions of condition $6^\bullet$. Since we assume that $6^\bullet$ holds, we obtain that
\begin{align*}
\forall \, (\varphi,\psi) \in \Psi(\mu,\nu) \quad \int\limits_{\M} \varphi\, d\mu - \int\limits_{\M} \psi\, d\nu \leq 0
\end{align*}
\noindent
and, therefore,
\begin{align*}
\sup\limits_{(\varphi,\psi) \in \Psi(\mu,\nu)} \left( \int\limits_{\M} \varphi\, d\mu - \int\limits_{\M} \psi\, d\nu \right) \leq 0.
\end{align*}
Using the~Kantorovich duality (\ref{duality}), we thus obtain that
\begin{align*}
\min\limits_{\pi \in \Pi(\mu,\nu)} \int\limits_{\M^2} c(p,q) \, d\pi(p,q) \leq 0.
\end{align*}
In particular, there exists at least one $\omega \in \Pi(\mu,\nu)$ such that the~integral above is \emph{finite}. But, by the~very definition of the~function $c$, this is possible iff $\omega(\M^2 \setminus J^+) = 0$ or, equivalently, iff $\omega(J^+) = 1$. Thus, we have proven the~existence of a~measure $\omega$ with desired properties.
\\

($7^\bullet \Rightarrow 2^\bullet$) Let $f \in C_b(\M)$ be a~causal function. Because the~probability measures $\mu$ and $\nu$ are, respectively, left and right marginals of the~joint distribution $\omega$, one can write that
\begin{align*}
\int\limits_{\M} f(p) d\mu(p) & = \int\limits_{\M^2} f(p) d\omega(p,q) = \int\limits_{J^+} f(p) d\omega(p,q)
\\
& \leq \int\limits_{J^+} f(q) d\omega(p,q) = \int\limits_{\M^2} f(q) d\omega(p,q) = \int\limits_{\M} f(q) d\nu(q),
\end{align*}
\noindent
where the~inequality follows from the~causality of $f$. In the~integrals with respect to $\omega$ we can always switch between $\M^2$ and $J^+$ because $\omega(\M^2 \setminus J^+) = 1 - \omega(J^+) = 0$.
\end{Proof}

The~fourth result concerns globally hyperbolic spacetimes. It provides an~additional characterisation of causality in terms of Cauchy hypersurfaces.
\begin{Thm}
\label{main_result4}
Let $\M$ be a~globally hyperbolic spacetime. Conditions $1^\bullet$--$7^\bullet$ are equivalent to the~following condition
\begin{enumerate}[1{$^\bullet$}]
\setcounter{enumi}{7}
\item For every Cauchy hypersurface\footnote{This includes nonsmooth and nonspacelike ones (considered Cauchy surfaces must be achronal, but need not be acausal).} ${\cal S} \subseteq \M$
    \begin{align}
    \label{causal8}
    \mu(J^+({\cal S})) \leq \nu(J^+({\cal S})) \, .
    \end{align}
\end{enumerate}
\end{Thm}
\begin{Proof}\textbf{:}
($5^\bullet \Rightarrow 8^\bullet$) Trivial.
\\

($8^\bullet \Rightarrow 4^\bullet$) Let ${\cal T}: \M \rightarrow \mathbb{R}$ be a~smooth temporal function whose every level set is a~Cauchy hypersurface.


Take any compact subset $\K \subseteq \M$. Let $T_0$ denote the~minimal value attained at $\K$ by the~function ${\cal T}$. For any $n \in \mathbb{N}_0$ define the~level set ${\cal S}_n := {\cal T}^{-1}(T_0 + n)$. Every ${\cal S}_n$ is a~smooth spacelike Cauchy hypersurface. Now, for any $n \in \mathbb{N}_0$ consider the~set
\begin{align*}
\Sigma_n := \partial J^+\left({\cal S}_n \cup \K\right).
\end{align*}

\begin{figure}[h]
\begin{center}
\psfrag{K}{\footnotesize $\K$}
\psfrag{JpK}{\footnotesize$J^+(\K)$}
\psfrag{s0}{\footnotesize${\cal S}_0 = \Sigma_0$}
\psfrag{s1}{\footnotesize${\cal S}_1$}
\psfrag{s2}{\footnotesize${\cal S}_2$}
\psfrag{si1}{\footnotesize$\Sigma_1$}
\psfrag{si2}{\footnotesize$\Sigma_2$}
\resizebox{!}{150pt}{\includegraphics[scale=0.6]{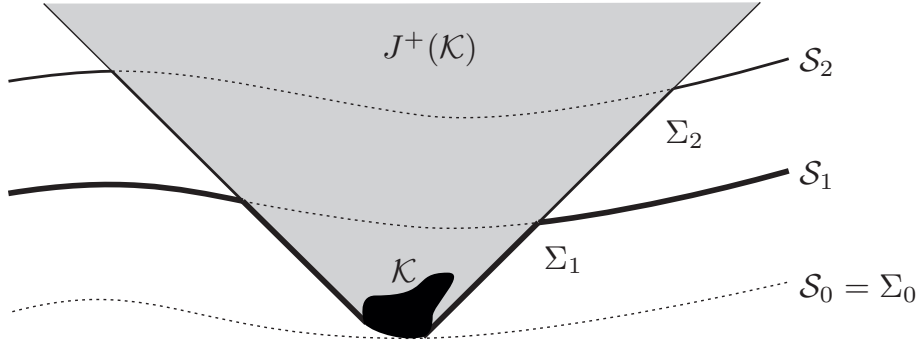}}
\caption{The construction of $\Sigma_n$'s.}
\end{center}
\end{figure}

We claim that for every $n \in \mathbb{N}_0$, $\Sigma_n$ is a~Cauchy hypersurface and that
\begin{align}
\label{compact}
J^+(\Sigma_n) = J^+\left({\cal S}_n \cup \K\right).
\end{align}
Indeed, observe first that $J^+\left({\cal S}_n \cup \K\right)$ is a~future set. By \cite[Chapter 14, Corollary 27]{BN83} $\Sigma_n$ is therefore a~closed achronal topological hypersurface. Let $\gamma$ be any inextendible timelike curve. It crosses the~Cauchy hypersurfaces ${\cal S}_n$ (which is contained in $J^+\left({\cal S}_n \cup \K\right)$) and ${\cal S}_0$ (the past of which, $I^-({\cal S}_0)$, is disjoint with $J^+\left({\cal S}_n \cup \K\right)$), therefore it must cross the~boundary $\partial J^+\left({\cal S}_n \cup \K\right) = \Sigma_n$. Since the~latter is achronal, it is met by $\gamma$ exactly once and therefore $\Sigma_n$ is a~Cauchy hypersurface.

In order to obtain (\ref{compact}), we prove the~following lemma.
\begin{Lem}
\label{lemat}
Let $\M$ be a~spacetime and let ${\cal F} \subseteq \M$ be a~closed future set such that ${\cal F} \subseteq J^+({\cal X})$ for some achronal set ${\cal X}$. Then $J^+(\partial {\cal F}) = {\cal F}$.
\end{Lem}
\begin{Proof}\textbf{:}
`$\subseteq$' Because ${\cal F}$ is closed, it contains its boundary: $\partial {\cal F} \subseteq {\cal F}$. Hence
\begin{align*}
J^+(\partial {\cal F}) \subseteq J^+({\cal F}) = {\cal F},
\end{align*}
\noindent
because ${\cal F}$ is a~future set.

`$\supseteq$' Take $q \in {\cal F}$. By assumption, there exists $x \in {\cal X}$ and a~future-directed causal curve $\gamma$ connecting $x$ with $q$.

Notice, first, that $x \not\in {\cal F} \setminus \partial {\cal F} = \textrm{int} \, {\cal F}$. Indeed, if $x$ would belong to $\textrm{int} \, {\cal F}$, which is an~open subset of ${\cal F}$, there would exist $x^\prime \in {\cal F}$ such that $x^\prime \ll x$. But since ${\cal F} \subseteq J^+({\cal X})$, there would exist $x^{\prime\prime} \in {\cal X}$ such that $x^{\prime\prime} \preceq x^{\prime}$. Altogether, by (\ref{prelim1}) we would obtain that $x^{\prime\prime} \ll x$, in contradiction with the~achronality of ${\cal X}$. Therefore, either $x \in \partial {\cal F}$ or $x \in \M \setminus {\cal F}$.

If $x \in \partial {\cal F}$, then $q \in J^+(\partial {\cal F})$ and the~proof is complete.

On the~other hand, if $x \in \M \setminus {\cal F}$, then the~curve $\gamma$ must cross $\partial {\cal F}$ at some point $p$. Of course, $p \preceq q$ and hence also in this case $q \in J^+(\partial {\cal F})$.
\end{Proof}

Notice now that $J^+\left({\cal S}_n \cup \K\right) = J^+({\cal S}_n) \cup J^+(\K)$ is in fact a~\emph{closed}\footnote{For the~closedness of $J^+({\cal S}_n)$, we refer e.g. to \cite[Section 10.2.7]{HR09}. The~closedness of $J^+(\K)$, on the~other hand, follows from the~causal simplicity of $\M$.} future set such that $J^+\left({\cal S}_n \cup K\right) \subseteq J^+({\cal S}_0)$. On the~strength of Lemma \ref{lemat}, we obtain (\ref{compact}).

By $8^\bullet$, because $\Sigma_n$ is a~Cauchy hypersurface for any $n \in \mathbb{N}_0$, we can write that
\begin{align}
\label{compact1}
\mu\left(J^+(\Sigma_n)\right) \leq \nu\left(J^+(\Sigma_n)\right).
\end{align}

Observe that the~sequence $\left( J^+(\Sigma_n) \right)_{n \in \mathbb{N}_0}$ is decreasing, because for all $n \in \mathbb{N}_0$
\begin{align*}
J^+(\Sigma_{n+1}) & = J^+\left({\cal S}_{n+1} \cup \K\right) = J^+({\cal S}_{n+1}) \cup J^+(\K)
\\
& = {\cal T}^{-1}\left( [T_0 + n + 1, +\infty) \right) \cup J^+(\K) \subseteq {\cal T}^{-1}\left( [T_0 + n, +\infty) \right) \cup J^+(\K) = J^+(\Sigma_n),
\end{align*}
\noindent
where we have used (\ref{compact}) and the~very definition of ${\cal S}_n$'s. Property (\ref{meas2}) allows us to pass with $n \rightarrow +\infty$ in (\ref{compact1}) and write that
\begin{align}
\label{compact2}
\mu\left( \bigcap\limits_{n = 0}^{\infty} J^+(\Sigma_n) \right) \leq \nu\left( \bigcap\limits_{n = 0}^{\infty} J^+(\Sigma_n) \right).
\end{align}
The~countable intersection appearing above can be easily shown to be equal to $J^+(\K)$. Indeed, one has
\begin{align*}
\bigcap\limits_{n = 0}^{\infty} J^+(\Sigma_n) & = J^+(\K) \cup \bigcap\limits_{n = 0}^{\infty} J^+({\cal S}_n) = J^+(\K) \cup \bigcap\limits_{n = 0}^{\infty} {\cal T}^{-1}\left( [T_0 + n, +\infty) \right)
\\
& = J^+(\K) \cup {\cal T}^{-1}\Big( \underbrace{\bigcap\limits_{n = 0}^{\infty} [T_0 + n, +\infty)}_{= \ \emptyset} \Big) = J^+(\K).
\end{align*}
Therefore, (\ref{compact2}) yields (\ref{causal4}) and the~proof of $4^\bullet$ is complete.
\end{Proof}

We have thus provided 8 different characterisations of a~causal relation between probability measures, which are equivalent if the~underlying spacetime is globally hyperbolic. Some of the~implications hold under lower causality conditions, as demonstrated in Theorems \ref{main_result1} -- \ref{main_result3}. Let us now discuss other implications not covered in the~proofs.

\begin{Rem}
\label{rem3}
Let us first stress that the~formulation of conditions $3^\bullet$--$5^\bullet$ using the~future of a~set is just a~matter of convention and one could equally well employ the~pasts. Concretely, straightforward application of the time inversion (note that such operation changes the~relation $\preceq$ into the~opposite one) shows that conditions $3^\bullet, 4^\bullet, 5^\bullet$ are (in any spacetime $\M$) equivalent to the~following conditions, respectively:
\begin{enumerate}[1{$^{\prime \bullet}$}]
\setcounter{enumi}{2}
\item For every open past set ${\cal P} \subseteq \M$
\begin{align}
    \label{causal3prime}
    \mu({\cal P}) \geq \nu({\cal P}) \, .
\end{align}
\item For every compact $\K \subseteq \M$
    \begin{align}
    \label{causal4prime}
    \mu(J^-(\K)) \geq \nu(J^-(\K)) \, .
    \end{align}
\item For every Borel past set ${\cal P} \subseteq \M$
    \begin{align}
    \label{causal5prime}
    \mu({\cal P}) \geq \nu({\cal P}) \, .
    \end{align}
\end{enumerate}
\end{Rem}

\begin{Rem}
\label{rem1}
Clearly, the~proof of the~implication $7^\bullet \Rightarrow 2^\bullet$ uses neither the~causal simplicity of $\M$ nor the~boundedness of the~function $f$. In fact, it works for any spacetime and for any $\mu$- and $\nu$-integrable causal function. We can, therefore, write down the~following condition
\begin{enumerate}[1{$^{\prime \bullet}$}]
\setcounter{enumi}{1}
\item For every causal $f \in \cL^1(\M, \mu) \cap \cL^1(\M, \nu)$,
    \begin{align}
    \label{causalprime}
    \int\limits_{\M} f \, d\mu \leq \int\limits_{\M} f \, d\nu.
    \end{align}
\end{enumerate}
For any spacetime $\M$ it is then true that $7^\bullet \Rightarrow 2^{\prime \bullet}$ as well as, trivially, $2^{\prime \bullet} \Rightarrow 2^\bullet \Rightarrow 1^\bullet$.
\end{Rem}

\begin{Rem}
\label{rem2}
Condition $2^{\prime \bullet}$ implies $5^\bullet$ in any spacetime $\M$.
\end{Rem}
\begin{Proof}\textbf{:}
Let ${\cal F}$ be a~Borel future subset of $\M$. Clearly, $\jdn_{\cal F} \in \cL^1(\M,\mu) \cap \cL^1(\M,\nu)$ and, by Corollary \ref{futurepastsets}, it is a~causal function. By condition $2^{\prime \bullet}$ we can write
\begin{align*}
\mu({\cal F}) = \int\limits_{\M} \jdn_{\cal F} d\mu \leq \int\limits_{\M} \jdn_{\cal F} d\nu = \nu({\cal F}),
\end{align*}
\noindent
what proves $5^\bullet$.
\end{Proof}

\begin{Rem}
\label{rem4}
Also the~implication $7^\bullet \Rightarrow 6^\bullet$ holds in all spacetimes. We can show even slightly more, namely, that condition $7^\bullet$ implies
\begin{enumerate}[1{$^{\prime \bullet}$}]
\setcounter{enumi}{5}
\item For all $\varphi, \psi: \M \rightarrow \mathbb{R}$ such that $\varphi$ is $\mu$-integrable and $\psi$ is $\nu$-integrable
    \begin{align}
    \label{Kantorovichprime}
    \big[ \forall p,q \in \M \quad p \preceq q \ \Rightarrow \ \varphi(p) \leq \psi(q) \big] \ \quad \Rightarrow \quad \ \int\limits_{\M} \varphi \, d\mu \leq \int\limits_{\M} \psi \, d\nu.
    \end{align}
\end{enumerate}
\end{Rem}
\begin{Proof}\textbf{:}
Similarly as in the~proof of $7^\bullet \Rightarrow 2^\bullet$, one can write that
\begin{align*}
\int\limits_{\M} \varphi(p) d\mu(p) & = \int\limits_{\M^2} \varphi(p) d\omega(p,q) = \int\limits_{J^+} \varphi(p) d\omega(p,q)
\\
& \leq \int\limits_{J^+} \psi(q) d\omega(p,q) = \int\limits_{\M^2} \psi(q) d\omega(p,q) = \int\limits_{\M} \psi(q) d\nu(q),
\end{align*}
\noindent
where the~inequality follows from the~assumptions on $\varphi$ and $\psi$.
\end{Proof}

\subsection{Basic properties of the~causal relation between measures}\label{subsec:omega}

In the~previous subsection we have shown that for \emph{any} spacetime $\M$ the~condition $7^\bullet$ not only implies all of the others listed in Theorems \ref{main_result1}, \ref{main_result2}, \ref{main_result3} and \ref{main_result4}, but also more general ones $2'^\bullet$ and $6'^\bullet$. It encourages us to promote the~condition $7^\bullet$ to a~\emph{definition} of the~causal precedence relation on $\mathfrak{P}(\M)$ for any spacetime $\M$.
\begin{Def}
\label{causality_def_true}
Let $\M$ be a~spacetime. For any $\mu,\nu \in \mathfrak{P}(\M)$ we say that $\mu$ \emph{causally precedes} $\nu$ (symbolically $\mu \preceq \nu$) iff there exists $\omega \in \mathfrak{P}(\M^2)$ such that
\begin{enumerate}[i)]
\item $(\pr_1)_\ast \omega = \mu$ and $(\pr_2)_\ast \omega = \nu$,
\item $\omega(J^+) = 1$.
\end{enumerate}
Such an~$\omega$ will be called a~\emph{causal coupling} of $\mu$ and $\nu$.
\end{Def}

Observe that $\omega(J^+)$ is well-defined because, by Theorem \ref{mainThm}, $J^+$ is $\sigma$-compact, and hence Borel, for any spacetime $\M$.

\begin{Rem}
\label{remomega}
In the~case of causally simple spacetimes $J^+ \subseteq \M^2$ is closed and therefore, by the~very definition of the~support of a~measure (see the~last paragraph of Section \ref{subsec:measure}), condition \emph{ii)} in Definition \ref{causality_def_true} is equivalent to the~inclusion $\supp \, \omega \subseteq J^+$. However, without the assumption of causal simplicity this is no longer true.
\end{Rem}

The~term `coupling (of measures $\mu$ and $\nu$)' comes from the~optimal transport theory \cite{CV03}, where it describes any $\omega \in \mathfrak{P}(\M^2)$ with property \emph{i)} of the~above definition. The~set of such couplings, denoted $\Pi(\mu, \nu)$, has already appeared above in the~context of the~Kantorovich duality (Theorem \ref{Kantorovich_duality}).

Such a~coupling --- or a~\emph{transference plan}, as it is also called --- can be regarded as an~instruction how to `reconfigure' a~fixed amount of `mass' distributed over $\M$ according to the~measure $\mu$ so that it becomes distributed according to the~measure $\nu$. This `reconfiguration' involves transporting the~(possibly infinitesimal) portions of `mass' between points of $\M$, and a~coupling $\omega \in \Pi(\mu,\nu) \subseteq \mathfrak{P}(\M^2)$ precisely describes what amount of `mass' is transported between any given pair of points.

It is, however, property \emph{ii)} which ties the~above definition with the~causality theory. It can be summarised as a~requirement that the~transport of `mass' be conducted along future-directed causal curves only --- that is why such couplings deserve to be called \emph{causal}. The~set of all causal couplings of measures $\mu$ and $\nu$ will be denoted by $\Pi_c(\mu,\nu)$.

Notice that a~(causal) coupling does \emph{not} specify along \emph{which} (causal) curves the~portions of `mass' are transported. In fact, various families of (causal) curves can lead to the~same (causal) coupling. Notice also that the~`mass' concentrated initially at some point $p \in \M$ can dilute to many different points.

Observe that for Dirac measures $\mu = \delta_p$, $\nu = \delta_q$ Definition \ref{causality_def_true} reduces to the~standard definition of the~causal relation between events $p$ and $q$. This can be seen as a~corollary of the~following proposition.
\begin{Prop}
\label{couplings}
Let $\M$ be a~topological space and let $\mu,\nu \in \mathfrak{P}(\M)$ and $\omega \in \Pi(\mu,\nu)$. Then, for any Borel sets $A,B \subseteq \M$
\begin{enumerate}[i)]
\item $\mu(A) = \nu(B) = 1 \quad \Leftrightarrow \quad \omega(A \times B) = 1$,
\item $\mu(A) = 0 \ \vee \ \nu(B) = 0 \quad \Rightarrow \quad \omega(A \times B) = 0$.
\end{enumerate}
\end{Prop}
\begin{Proof}\textbf{:}
i) To prove `$\Rightarrow$' we use the~inclusion--exclusion principle to write
\begin{align*}
& 1 \geq \omega(A \times B) = \omega(A \times \M \, \cap \, \M \times B) = \underbrace{\omega(A \times \M)}_{= \, \mu(A) \, = \, 1} + \underbrace{\omega(\M \times B)}_{= \, \nu(B) \, = \, 1} - \underbrace{\omega(A \times \M \, \cup \, \M \times B)}_{\leq \, 1}
\\
& \geq 1 + 1 - 1 = 1.
\end{align*}
Conversely, to prove `$\Leftarrow$', notice that
\begin{align*}
1 \geq \mu(A) = \omega(A \times \M) \geq \omega(A \times B) = 1 \quad \textnormal{and} \quad 1 \geq \nu(B) = \omega(\M \times B) \geq \omega(A \times B) = 1.
\end{align*}

ii) One has
\begin{align*}
\vspace{-0.5cm}
0 \leq \omega(A \times B) \leq \min \left\{ \omega(A \times \M), \omega(\M \times B) \right\} = \min \left\{ \mu(A), \nu(B) \right\} = 0.
\end{align*}
\vspace*{-0.5cm}
\end{Proof}

\begin{Cor}
\label{couplingsCor1}
Let $\M$ be a~spacetime. Then for any $p,q \in \M$ $p \preceq q$ iff $\delta_p \preceq \delta_q$.
\end{Cor}
\begin{Proof}\textbf{:}
By Proposition \ref{couplings}, the~only coupling between two Dirac measures $\delta_p, \delta_q$ is their product measure $\omega := \delta_p \times \delta_q = \delta_{(p,q)}$. Hence, the~fact that $p \preceq q$ is equivalent in this case to the~requirement that $\omega(J^+) = 1$.
\end{Proof}
\vspace*{-0.2cm}
\begin{Cor}
\label{couplingsCor2}
Let $\M$ be a~causally simple spacetime. For any $p,q \in \M$ the~following conditions are equivalent
\begin{enumerate}[1{$^\diamond$}]
\item $\forall \, f \in {\cal C}(\M) \quad f(p) \leq f(q)$,
\item $p \preceq q$.
\end{enumerate}
\end{Cor}
\begin{Proof}\textbf{:}
It is a~direct consequence of the~equivalence ($1^\bullet \Rightarrow 7^\bullet$) in Theorem \ref{main_result3} and Corollary \ref{couplingsCor1}.
\end{Proof}

If the~measure $\mu$ is compactly supported, then in the~light of the~above discussion it is natural to expect that the~support of any $\nu$ with $\mu \preceq \nu$ should be within the~future of $\supp \, \mu$ \cite{WSWSG11}. This intuitive condition in fact true in causally simple spacetimes.
\begin{Prop}
\label{neccond}
Let $\M$ be a~spacetime and let $\mu, \nu \in \mathfrak{P}(\M)$, with $\mu$ compactly supported and $\mu \preceq \nu$. Then $\nu(J^+(\supp \, \mu)) = 1$. Moreover, if $\M$ is causally simple then $\supp \, \nu \subseteq J^+(\supp \, \mu)$.
\end{Prop}
\begin{Proof}\textbf{:}
By condition $4^\bullet$ (which is implied by Definition \ref{causality_def_true}) it is true that
\begin{align*}
1 = \mu(\supp \, \mu) \leq \mu(J^+(\supp \, \mu)) \leq \nu(J^+(\supp \, \mu)) \leq 1,
\end{align*}
and therefore $\nu(J^+(\supp \, \mu)) = 1$.

We now claim that if $\M$ is causally simple, then this implies that $\supp \, \nu \subseteq J^+(\supp \, \mu)$.

Indeed, recall that in a~causally simple spacetime the~causal futures of compact sets are closed. Therefore, if there existed $q \in \supp \, \nu$ but $q \not\in J^+(\supp \, \mu)$, then we could take an~open neighborhood $U \ni q$ such that $\nu(U) > 0$ but $U \cap J^+(\supp \, \mu) = \emptyset$. But this would imply that \vspace{-0.2cm}
\begin{align*}
\nu(J^+(\supp \, \mu)) \leq 1 - \nu(U) < 1,
\end{align*}
in contradiction with the~first part of the~proof.
\end{Proof}
\pagebreak


Recall that the~causal precedence relation between events is reflexive, transitive and, iff $\M$ is causal, antisymmetric. We now prove analogous results for the~space of Borel probability measures on $\M$ equipped with the~relation $\preceq$. To this end, it will be convenient to use the~\emph{diagonal function} $\Delta: \M \rightarrow \M^2$, defined as $\Delta(p) := (p,p)$ for any $p \in \M$.

\begin{Thm}
Let $\M$ be a~spacetime. The~relation $\preceq$ on $\mathfrak{P}(\M)$ is reflexive and transitive.
\end{Thm}
\begin{Proof}\textbf{:}
To prove reflexivity of $\preceq$, it suffices to notice that for any $\mu \in \mathfrak{P}(\M)$ the~pushforward measure $\Delta_\ast \mu$ is a~causal coupling of $\mu$ with itself.

Indeed, $(\pr_i)_\ast \Delta_\ast \mu = \left( \pr_i \circ \Delta \right)_\ast \mu = \textnormal{id}_\ast \mu = \mu$ for $i=1,2$ and $\Delta_\ast \mu(J^+) = \mu(\Delta^{-1}(J^+)) = \mu(\M) = 1$, where we have used the~equality $\Delta^{-1}(J^+) = \M$, which expresses nothing but the~reflexivity of the~causal precedence relation between \emph{events}.

We now move to proving the~transitivity of $\preceq$. Let us invoke the~following standard result \cite[Lemma 7.6]{CV03} from the~optimal transport theory.
\begin{Lem}\textbf{(Gluing Lemma)}
\label{gluing}
Let $({\cal X}_i, \mu_i)$, $i=1,2,3$ be Polish probability spaces and assume there exist couplings $\omega_{12} \in \Pi(\mu_1,\mu_2)$ and $\omega_{23} \in \Pi(\mu_2,\mu_3)$.

Then, there exists $\omega_{123} \in \mathfrak{P}({\cal X}_1 \times {\cal X}_2 \times {\cal X}_3)$ such that $\left(\pr_{12}\right)_\ast \omega_{123} = \omega_{12}$ and $\left(\pr_{23}\right)_\ast \omega_{123} = \omega_{23}$, where $\pr_{ij}: {\cal X}_1 \times {\cal X}_2 \times {\cal X}_3 \rightarrow {\cal X}_i \times {\cal X}_j$ denotes the~canonical projection map.

Moreover, $\omega_{13} := \left(\pr_{13}\right)_\ast \omega_{123}$ belongs to $\Pi(\mu_1,\mu_3)$.
\end{Lem}
The~Gluing Lemma works well with the~causal precedence relation. Concretely, let us take $\mu_1,\mu_2,\mu_3 \in \mathfrak{P}({\cal M})$ such that $\mu_1 \preceq \mu_2 \preceq \mu_3$, where $\omega_{12} \in \Pi_c(\mu_1,\mu_2)$ and $\omega_{23} \in \Pi_c(\mu_2,\mu_3)$. Then the~coupling $\omega_{13}$ of $\mu_1$ and $\mu_3$ is causal, too.

Indeed, notice first that
\begin{multline*}
\omega_{123} \left( \left\{ (p,q,r) \in \M^3 \ | \ p \preceq q \not\preceq r \right\} \right) \leq \omega_{123} \left( \left\{ (p,q,r) \in \M^3 \ | \ q \not\preceq r \right\} \right)
\\
= \omega_{123}\left( \M \times \left(\M^2 \setminus J^+ \right) \right) = \omega_{23}\left(\M^2 \setminus J^+ \right) = 1 - \omega_{23}(J^+) = 0,
\end{multline*}
\noindent
and thus $\omega_{123} \left( \left\{ (p,q,r) \in \M^3 \ | \ p \preceq q \not\preceq r \right\} \right) = 0$.

On the other hand,
\begin{multline*}
\omega_{123} \left( \left\{ (p,q,r) \in \M^3 \ | \ p \not\preceq q \right\} \right) = \omega_{123}\left( \left(\M^2 \setminus J^+ \right) \times \M \right) = \omega_{12}\left(\M^2 \setminus J^+ \right)
\\
= 1 - \omega_{12}(J^+) = 0.
\end{multline*}

Since $\M^3$ can be decomposed into the~following union of (pairwise disjoint) sets
\begin{align*}
\M^3 & = \left\{ (p,q,r) \in \M^3 \ | \ p \preceq q \preceq r \right\} \, \cup \, \left\{ (p,q,r) \in \M^3 \ | \ p \preceq q \not\preceq r \right\}
\\
& \cup \, \left\{ (p,q,r) \in \M^3 \ | \ p \not\preceq q \right\},
\end{align*}
\noindent
therefore we obtain
\begin{align*}
1 & = \omega_{123}(\M^3) = \omega_{123} \left( \left\{ (p,q,r) \in \M^3 \ | \ p \preceq q \preceq r \right\} \right)
\\
& + \underbrace{\omega_{123} \left( \left\{ (p,q,r) \in \M^3 \ | \ p \preceq q \not\preceq r \right\} \right)}_{= \, 0} + \underbrace{\omega_{123} \left( \left\{ (p,q,r) \in \M^3 \ | \ p \not\preceq q \right\} \right)}_{= \, 0}
\end{align*}
\noindent
and hence
\begin{align}
\label{gluing1}
\omega_{123} \left( \left\{ (p,q,r) \in \M^3 \ | \ p \preceq q \preceq r \right\} \right) = 1.
\end{align}

But this, in turn, means that
\begin{align*}
1 \geq \omega_{13}(J^+) = \omega_{123}\left( \left\{ (p,q,r) \in \M^3 \ | \ p \preceq r \right\} \right) \geq \omega_{123}\left( \left\{ (p,q,r) \in \M^3 \ | \ p \preceq q \preceq r \right\} \right) = 1,
\end{align*}
\noindent
where the~middle inequality is a~direct consequence of the~transitivity of the~causal precedence relation between events. We have thus proven that $\omega_{13}(J^+) = 1$, and so $\omega_{13} \in \Pi_c(\mu_1,\mu_3)$ and therefore $\mu_1 \preceq \mu_3$.
\end{Proof}

The~natural question arises: how robust the causal structure of a spacetime ${\cal M}$ must be to render the~relation $\preceq$ antisymmetric and hence a~partial order? Obviously, ${\cal M}$ must be at least causal (otherwise even the~causal precedence relation between \emph{events} fails to be antisymmetric).

We have the~following partial result.
\begin{Thm}
\label{antisymmetry_partial}
Let ${\cal M}$ be a~spacetime with the~following property:

For any compact $\K \subseteq \M$ there exists a~Borel function $\tau_\K: \K \rightarrow \mathbb{R}$ such that
\begin{align}
\label{propertyP}
\forall \, p,q \in \K \quad p \preceq q \ \Rightarrow\  \tau_\K(p) < \tau_\K(q).
\end{align}
\noindent
Then, for any $\mu \in \mathfrak{P}(\M) \, \ \Pi_c(\mu, \mu) = \{ \Delta_\ast \mu \}$. Moreover, the~relation $\preceq$ is antisymmetric.
\end{Thm}

\begin{Rem}
\label{property}
Property (\ref{propertyP}) implies that $\M$ is causal. Indeed, suppose that there exist two distinct events $p,q \in \M$ such that $p \preceq q \preceq p$. Taking now $\K = \{p,q\}$, on the~strength of (\ref{propertyP}) we would obtain that $\tau_\K(p) < \tau_\K(q) < \tau_\K(p)$, a~contradiction.

On the~other hand, if $\M$ is past (future) distinguishing, then any past (resp. future) volume function is a~semi-continuous, and hence Borel, generalised time function (cf. Section \ref{subsec:causality}). This obviously implies (\ref{propertyP}) --- for any compact $\K \subseteq \M$ simply define $\tau_\K := \tau|_\K$. However, being past or future distinguishing is not necessary for (\ref{propertyP}) to hold. Indeed, the~rightmost diagram in \cite[Figure 6]{MS08} presents a~causal, but neither future nor past distinguishing spacetime $\M := \mathbb{R} \times S^1 \setminus \{(0,0)\}$, which admits a~Borel generalised time function, for instance
\begin{align*}
\tau(x, \theta) := \left\{ \begin{array}{ll}
                           \arctan{x} & \textnormal{for } x < 0 \\
                           \theta & \textnormal{for } x = 0 \\
                           2\pi + \arctan{x} & \textnormal{for } x > 0
                           \end{array} \right. \, ,
\end{align*}
\noindent
for any $x \in \mathbb{R}$ and $\theta \in S^1$, where the~latter is the~angular coordinate whose range is $[0,2\pi)$, except for $x = 0$, when its range is $(0,2\pi)$.
\end{Rem}

Before we move to the~proof of Theorem \ref{antisymmetry_partial}, let us present the~following lemma.
\begin{Lem}
\label{measurelemma}
Let $\M$ be a~topological space and let $\mu,\nu$ be two Borel probability measures on $\M$. Finally, let $\omega \in \Pi(\mu,\nu)$ be such that $\omega(\Delta(\M)) = 1$. Then $\mu = \nu$ and $\omega = \Delta_\ast \mu = \Delta_\ast \nu$.
\end{Lem}
\begin{Proof}\textbf{:}
Let ${\cal U}$ be any Borel subset of $\M^2$. Then, $\omega({\cal U} \setminus \Delta(\M)) \leq \omega(\M \setminus \Delta(\M)) = 1 - \omega(\Delta(\M)) = 0$ and therefore $\omega({\cal U} \setminus \Delta(\M)) = 0$. But this allows us to write
\begin{align*}
\omega({\cal U}) & = \omega({\cal U} \cap \Delta(\M)) + \underbrace{\omega({\cal U} \setminus \Delta(\M))}_{= \, 0} = \omega\left( \Delta(\Delta^{-1}({\cal U})) \right).
\end{align*}
The~rightmost expression, in turn, can be further transformed either into
\begin{align*}
& \omega\left( \Delta(\Delta^{-1}({\cal U})) \right) = \omega \left( (\Delta^{-1}({\cal U}) \times \M) \cap \Delta(\M) \right)
\\
& = \omega \left( \Delta^{-1}({\cal U}) \times \M \right) - \underbrace{\omega\left( (\Delta^{-1}({\cal U}) \times \M) \setminus \Delta(\M) \right)}_{= \, 0} = \mu\left( \Delta^{-1}({\cal U}) \right) = \Delta_\ast \mu ({\cal U})
\end{align*}
\noindent
or into
\begin{align*}
& \omega\left( \Delta(\Delta^{-1}({\cal U})) \right) = \omega \left( (\M \times \Delta^{-1}({\cal U})) \cap \Delta(\M) \right)
\\
& = \omega \left( \M \times \Delta^{-1}({\cal U}) \right) - \underbrace{\omega\left( \M \times (\Delta^{-1}({\cal U}) ) \setminus \Delta(\M) \right)}_{= \, 0} = \nu\left( \Delta^{-1}({\cal U}) \right) = \Delta_\ast \nu ({\cal U}),
\end{align*}
\noindent
what proves the~second part of the~theorem. To obtain the~equality $\mu = \nu$, take any Borel ${\cal V} \subseteq \M$ and notice, for instance, that
\begin{align*}
\nu({\cal V}) = \omega(\M \times {\cal V}) = \Delta_\ast \mu(\M \times {\cal V}) = \mu\left( \Delta^{-1}(\M \times {\cal V}) \right) = \mu({\cal V}),
\end{align*}
which concludes the~entire proof.
\end{Proof}

\begin{Proof} \textbf{of Theorem \ref{antisymmetry_partial}:}
Take any $\mu \in \mathfrak{P}(\M)$ and let $\pi \in \Pi_c(\mu,\mu)$. By Definition \ref{causality_def_true}, we have that
\begin{align*}
\forall \, f \in {\cal L}^1(\M, \mu) \quad \int\limits_{J^+} f(p) d\pi(p,q) = \int\limits_{\M} f \, d\mu = \int\limits_{J^+} f(q) d\pi(p,q)
\end{align*}
\noindent
and hence
\begin{align*}
\forall \, f \in {\cal L}^1(\M, \mu) \quad \int\limits_{J^+} \left(f(q) - f(p)\right) d\pi(p,q) = 0
\end{align*}
\noindent
or, by noticing that the~integrand vanishes on $\Delta(\M)$,
\begin{align}
\label{antisymmetry1}
\forall \, f \in {\cal L}^1(\M, \mu) \quad \int\limits_{J^+ \setminus \Delta(\M)} \left(f(q) - f(p)\right) d\pi(p,q) = 0.
\end{align}

Suppose now that $\pi(J^+ \setminus \Delta(\M)) > 0$. Because $\pi$ is tight, there exists a~compact set $K \subseteq J^+ \setminus \Delta(\M)$ with $\pi(K) > 0$. Notice that $K \subseteq \K^2$, where $\K := \pr_1 K \cup \pr_2 K$ is a~compact subset of $\M$, and so $\pi(\K^2 \cap J^+ \setminus \Delta(\M)) > 0$. Define $f_\K: \M \rightarrow \mathbb{R}$ via
\begin{align*}
f_\K(p) := \left\{ \begin{array}{ll} \arctan \tau_\K(p) & \textnormal{for } p \in \K \\ 0 & \textnormal{for } p \not\in \K \end{array} \right. ,
\end{align*}
\noindent
where $\tau_\K$ is a~function whose existence is guaranteed by property (\ref{propertyP}). Function $f_\K$ is Borel and bounded, and hence $\mu$-integrable. Plugging it into (\ref{antisymmetry1}) yields
\begin{align*}
\int\limits_{\K^2 \cap J^+ \setminus \Delta(\M)} \left(\arctan \tau_\K(q) - \arctan \tau_\K(p)\right) d\pi(p,q) = 0.
\end{align*}
But the~integrand of the~above integral is positive on $\K^2 \cap J^+ \setminus \Delta(\M)$ by the~very definition of $\tau_\K$, therefore the~fact that the~integral is zero implies that $\pi(\K^2 \cap J^+ \setminus \Delta(\M)) = 0$, which contradicts the~earlier result. This proves that $\pi(J^+ \setminus \Delta(\M)) = 0$.

By property $ii)$ from Definition \ref{causality_def_true}, this in turn means that
\begin{align*}
\pi(\Delta(\M)) = \pi(J^+) - \pi(J^+ \setminus \Delta(\M)) = 1.
\end{align*}
On the~strength of Lemma \ref{measurelemma}, we get that $\pi = \Delta_\ast \mu$.
\\

We now move to proving the~antisymmetricity of the~relation $\preceq$. Let $\mu, \nu \in \mathfrak{P}(\M)$ be such that $\mu \preceq \nu \preceq \mu$. Let $\omega \in \Pi_c(\mu, \nu)$ and $\varpi \in \Pi_c(\nu, \mu)$. By the~Gluing Lemma, there exists $\Omega \in \mathfrak{P}(\M^3)$ such that $(\pr_{12})_\ast \Omega = \omega$, $(\pr_{23})_\ast \Omega = \varpi$ and $(\pr_{13})_\ast \Omega \in \Pi_c(\mu, \mu)$, which, by the~previous part of the~proof, means that $(\pr_{13})_\ast \Omega = \Delta_\ast \mu$.

Formula (\ref{gluing1}) takes here the~following form
\begin{align*}
\Omega \left( \left\{ (p,q,r) \in \M^3 \ | \ p \preceq q \preceq r \right\} \right) = 1.
\end{align*}

Notice, however, that the~set $\left\{ (p,q,r) \in \M^3 \ | \ p \preceq q \preceq r \neq p \right\}$ is $\Omega$-null, because
\begin{align*}
& \Omega \left( \left\{ (p,q,r) \in \M^3 \ | \ p \preceq q \preceq r \neq p \right\} \right) \leq \Omega \left( \left\{ (p,q,r) \in \M^3 \ | \ p \neq r \right\} \right)
\\
& = (\pr_{13})_\ast \Omega \left( \left\{ (p,r) \in \M^2 \ | \ p \neq r \right\} \right) = \Delta_\ast \mu \left( \M^2 \setminus \Delta(\M) \right) = 1 - \mu(\M) = 0.
\end{align*}
Therefore, in fact,
\begin{align}
\label{antisymmetric2}
& \Omega \left( \left\{ (p,q,p) \in \M^3 \ | \ p \preceq q \preceq p \right\} \right)
\\
\nonumber
& = \underbrace{\Omega \left( \left\{ (p,q,r) \in \M^3 \ | \ p \preceq q \preceq r \right\} \right)}_{= \, 1} - \underbrace{\Omega \left( \left\{ (p,q,r) \in \M^3 \ | \ p \preceq q \preceq r \neq p \right\} \right)}_{= \, 0} = 1.
\end{align}
But $\M$ is causal (cf. Remark \ref{property}), therefore the~causal precedence relation between events is antisymmetric and thus the~set whose measure is evaluated in (\ref{antisymmetric2}) is equal to $\{ (p,p,p) \in \M^3 \ | \ p \in \M \}$.

We can now easily obtain that
\begin{align*}
\omega(\Delta(\M)) & = \Omega(\Delta(\M) \times \M) = \Omega\left( \{ (p,p,q) \in \M^3 \ | \ p,q \in \M \} \right)
\\
& \geq \Omega \left( \{ (p,p,p) \in \M^3 \ | \ p \in \M \} \right) = 1
\end{align*}
\noindent
and so $\omega(\Delta(\M)) = 1$. Invoking Lemma \ref{measurelemma}, we obtain that $\mu = \nu$.
\end{Proof}

\section{Lorentz--Wasserstein distances}\label{sec:LW}

Recall that the~Lorentzian distance $d: \M^2 \rightarrow [0,+\infty]$ provides a~physically meaningful way of measuring distances between events, in an~analogy with the~Riemannian distance $d_R$ in the~case of Riemannian manifolds. In the~latter case, one can extend the~notion of a~distance to the space of measures on $\M$. Concretely, for any $s \geq 1$ one defines the~so-called \emph{$s^\textnormal{th}$ Wasserstein distance} between any two measures $\mu,\nu \in \mathfrak{P}({\cal R})$ on a~Riemannian manifold ${\cal R}$ as
\begin{align}
\label{Wasserstein}
W_s(\mu, \nu) := \inf\limits_{\pi \in \Pi(\mu,\nu)} \left[ \int\limits_{{\cal R}^2} d_R(x,y)^s d\pi(x,y) \right]^{1/s}.
\end{align}
For an~exposition of the~theory of Wasserstein distances in the~context of the~optimal transport theory one is referred e.g. to \cite{CV03}.

We now propose the~following natural definition of a~distance between measures on a~spacetime.

\begin{Def}
\label{LorentzWasserstein}
Let $\M$ be a~spacetime and let $s \in (0,1]$. The~\emph{$s^\textnormal{th}$ Lorentz--Wasserstein distance} is the~map $LW_s: \mathfrak{P}(\M) \times \mathfrak{P}(\M) \rightarrow [0,+\infty]$ given by
\begin{align*}
LW_s(\mu, \nu) := \left\{ \begin{array}{ll}
                         \sup\limits_{\omega \in \Pi_c(\mu,\nu)} \left[ \int\limits_{{\cal M}^2} d(p,q)^s d\omega(p,q) \right]^{1/s} & \textnormal{if } \Pi_c(\mu,\nu) \neq \emptyset
                         \\
                         0 & \textnormal{if } \Pi_c(\mu,\nu) = \emptyset
                         \end{array} \right. \cdot
\end{align*}
\end{Def}
Notice that the~integrals are well-defined, because $d$ is lower semi-continuous and hence Borel. Notice also that for Dirac measures $LW_s(\delta_p, \delta_q) = d(p,q)$ for any $s$.

Lorentz--Wasserstein distances have properties analogous to those of the~Lorentzian distance (cf. Section \ref{subsec:causality}).
\begin{Thm}
\label{LW0}
Let $\M$ be a~spacetime and let $s \in (0,1]$. Then:
\begin{enumerate}[i)]
\item For any $\mu,\nu \in \mathfrak{P}(\M)$
\begin{align*}
LW_s(\mu, \nu) > 0 \quad \Leftrightarrow \quad \exists \, \omega \in \Pi_c(\mu,\nu) \quad \omega(I^+) > 0 \quad \Rightarrow \quad \mu \preceq \nu.
\end{align*}
\item The~reverse triangle inequality holds. Namely, for any $\mu_1, \mu_2, \mu_3 \in \mathfrak{P}(\M)$
\begin{align}
\label{LWtriangle}
\mu_1 \preceq \mu_2 \preceq \mu_3 \quad \Rightarrow \quad LW_s(\mu_1, \mu_2) + LW_s(\mu_2, \mu_3) \leq LW_s(\mu_1, \mu_3).
\end{align}
\item For any $\mu \in \mathfrak{P}(\M)$, $LW_s(\mu, \mu)$ is either $0$ or $+\infty$.
\item $\M$ is chronological iff $\ \forall \, \mu \in \mathfrak{P}(\M) \quad LW_s(\mu, \mu) = 0$.
\item For any $\mu,\nu \in \mathfrak{P}(\M)$, if $LW_s(\mu, \nu) \in (0,+\infty)$ then $LW_s(\nu, \mu) = 0$.
\end{enumerate}
\end{Thm}
\begin{Proof}\textbf{:}
\emph{i)} The~implication is obvious, so we only prove the~equivalence.

To prove the~`$\Rightarrow$' part of the~equivalence, assume that $LW_s(\mu, \nu) > 0$. By the~very definition of $LW_s$, this implies that there exists $\omega \in \Pi_c(\mu, \nu)$ such that $\int\limits_{\M^2} d(p,q)^s d\omega(p,q) > 0$. In order to prove that $\omega(I^+) > 0$, suppose on the~contrary that $I^+$ is $\omega$-null. Then, one would have
\begin{align*}
0 & < \int\limits_{\M^2}\! d(p,q)^s d\omega(p,q) = \int\limits_{J^+}\! d(p,q)^s d\omega(p,q) = \!\!\!\!\!\! \underbrace{\int\limits_{E^+}\! d(p,q)^s d\omega(p,q)}_{= \, 0 \textnormal{, because } d \textnormal{ vanishes on } E^+} \!\! +\, \underbrace{\int\limits_{I^+}\! d(p,q)^s d\omega(p,q)}_{= \, 0 \textnormal{, because } \omega(I^+) \, = \, 0} = 0,
\end{align*}
\noindent
hence a~contradiction.

To prove the~`$\Leftarrow$' part, suppose there exists $\omega \in \Pi_c(\mu,\nu)$ with $\omega(I^+) > 0$, but nevertheless $LW_s(\mu, \nu) = 0$. The~latter implies that $\int\limits_{\M^2} d(p,q)^s d\omega(p,q) = 0$. But this, in turn, means that
\begin{align*}
\int\limits_{I^+} d(p,q)^s d\omega(p,q) = \int\limits_{J^+} d(p,q)^s d\omega(p,q) - \underbrace{\int\limits_{E^+} d(p,q)^s d\omega(p,q)}_{= \, 0 \textnormal{, because } d \textnormal{ vanishes on } E^+} = \int\limits_{\M^2} d(p,q)^s d\omega(p,q) = 0.
\end{align*}
But $d$ is positive on $I^+$ and so the~latter must be an~$\omega$-null set, which contradicts with the~assumption that $\omega(I^+) > 0$.
\\

\emph{ii)} 
Let $\mu_1, \mu_2, \mu_3 \in \mathfrak{P}(\M)$ satisfy $\mu_1 \preceq \mu_2 \preceq \mu_3$. Let $\omega_{12}$ and $\omega_{23}$ be any elements of $\Pi_c(\mu_1, \mu_2)$ and $\Pi_c(\mu_2, \mu_3)$, respectively, and let $\omega_{123} \in \mathfrak{P}(\M^3)$ be a~measure `gluing them together' as specified in the~Gluing Lemma. Recall from the~discussion following that lemma that $\omega_{13} := (\pr_{13})_\ast \omega_{123} \in \Pi_c(\mu_1, \mu_3)$.

One has the~inequality
\begin{align}
\label{LW1}
LW_s(\mu_1, \mu_3) \geq \left[ \int\limits_{\M^2} d(p,q)^s d\omega_{12}(p,q) \right]^{1/s} + \left[ \int\limits_{\M^2} d(q,r)^s d\omega_{23}(q,r) \right]^{1/s},
\end{align}
\noindent
which is proven through the~following sequence of equalities and inequalities.
\begin{align*}
LW_s(\mu_1, \mu_3) & \geq \left[ \int\limits_{\M^2} d(p,r)^s d\omega_{13}(p,r) \right]^{1/s} = \left[ \int\limits_{\M^3} d(p,r)^s d\omega_{123}(p,q,r) \right]^{1/s}
\\
& \geq \left[ \int\limits_{\M^3} \left( d(p,q) + d(q,r) \right)^s d\omega_{123}(p,q,r) \right]^{1/s}
\\
& \geq \left[ \int\limits_{\M^3} d(p,q)^s d\omega_{123}(p,q,r) \right]^{1/s} + \left[ \int\limits_{\M^3} d(q,r)^s d\omega_{123}(p,q,r) \right]^{1/s}
\\
& = \left[ \int\limits_{\M^2} d(p,q)^s d\omega_{12}(p,q) \right]^{1/s} + \left[ \int\limits_{\M^2} d(q,r)^s d\omega_{23}(q,r) \right]^{1/s},
\end{align*}
\noindent
where we have used, successively, the~definition of $LW_s$, the~Gluing Lemma (the definition of $\omega_{13}$), the~reverse triangle inequality for $d$, the~reverse Minkowski inequality for integrals \cite[Proposition 5.3.1]{garling2007inequalities} and, finally, the~Gluing Lemma again (the definition of $\omega_{123}$).

By the~arbitrariness of $\omega_{12} \in \Pi_c(\mu_1, \mu_2)$ and $\omega_{23} \in \Pi_c(\mu_2, \mu_3)$, inequality (\ref{LW1}) immediately yields (\ref{LWtriangle}) --- one simply has to take the~supremum over all $\omega_{12} \in \Pi_c(\mu_1, \mu_2)$ and all $\omega_{23} \in \Pi_c(\mu_2, \mu_3)$.
\\

\emph{iii)}
By \emph{ii)} and the~fact that $\mu \preceq \mu$, one has $2 LW_s(\mu, \mu) \leq LW_s(\mu, \mu)$, which is true iff either $LW_s(\mu, \mu) = 0$ or $LW_s(\mu, \mu) = +\infty$.
\\

\emph{iv)}
To prove `$\Rightarrow$', assume that $\M$ is chronological. By \emph{i)}, it suffices to show that for any $\mu \in \mathfrak{P}(\M)$ and for any $\omega \in \Pi_c(\mu,\mu)$ we must have $\omega(I^+) = 0$.

Indeed, proceeding identically as in the~beginning of the~proof of Theorem \ref{antisymmetry_partial}, we obtain (compare with (\ref{antisymmetry1}))
\begin{align}
\label{LW2}
\forall \, f \in {\cal L}^1(\M, \mu) \quad \int\limits_{J^+ \setminus \Delta(\M)} \left(f(q) - f(p)\right) d\omega(p,q) = 0.
\end{align}
The~key now is to use a~past volume function $t^-$ associated to some admissible measure on $\M$. Recall that $t^-$ is causal. Moreover, since $\M$ is chronological, $t^-$ is increasing on any future-directed timelike curve (cf. Section \ref{subsec:causality}). Symbolically:
\begin{align}
\label{LW3}
\forall \, (p,q) \in J^+ \quad t^-(p) \leq t^-(q) \qquad \textnormal{and} \qquad \forall \, (p,q) \in I^+ \quad t^-(p) < t^-(q).
\end{align}
Substituting $f := t^-$ in (\ref{LW2}) (recall that $t^-$ is Borel and bounded and hence $\mu$-integrable), we can write
\begin{align}
\label{LW4}
\int\limits_{E^+ \setminus \Delta(\M)} \left(t^-(q) - t^-(p)\right) d\omega(p,q) + \int\limits_{I^+} \left(t^-(q) - t^-(p)\right) d\omega(p,q) = 0.
\end{align}
By the~first property in (\ref{LW3}), both integrals in (\ref{LW4}) are nonnegative and hence they both must vanish. However, by the~second property in (\ref{LW3}), the~integrand in the~rightmost integral is positive on $I^+$, therefore this integral cannot vanish unless $\omega(I^+) = 0$.

The~proof of `$\Leftarrow$' is straightforward. Take any $p \in \M$ and notice that, by assumption,
\begin{align*}
d(p,p) = LW_s(\delta_p, \delta_p) = 0.
\end{align*}
But this implies (see property i) of the~Lorentzian distance in Section \ref{subsec:causality}) that $p \not\ll p$ for any $p \in \M$, which means that $\M$ is chronological.
\\

\emph{v)}
Suppose that $LW_s(\mu,\nu) \in (0,+\infty)$ but, nevertheless, $LW_s(\nu,\mu) > 0$. By \emph{i)}, this implies that $\mu \preceq \nu \preceq \mu$. By \emph{ii)}, we can write that
\begin{align*}
0 < LW_s(\mu,\nu) + LW_s(\nu,\mu) \leq LW_s(\mu,\mu).
\end{align*}
On the~other hand, again by \emph{ii)}, it is also true that
\begin{align*}
LW_s(\mu,\mu) + LW_s(\mu,\nu) \leq LW_s(\mu,\nu),
\end{align*}
\noindent
which, since $LW_s(\mu,\nu)$ is assumed finite, implies that $LW_s(\mu,\mu) \leq 0$ and we have arrived to a~contradiction.
\end{Proof}

Unlike the~Lorentzian distance, Lorentz--Wasserstein distances can assume infinite values even in globally hyperbolic spacetimes.
\begin{Ex}
Consider the~$(1+1)$-dimensional Minkowski spacetime $\M := \mathbb{R}^{1,1}$ and fix $s \in (0,1]$. Let $\mu := \delta_{(0,0)}$ and $\nu := \sum\limits_{i=1}^{\infty} 2^{-i} \delta_{(2^{i/s},0)}$. Define $\omega \in \Pi_c(\mu,\nu)$ by\footnote{In fact, it is the~only causal coupling between those particular $\mu$ and $\nu$.}
\begin{align*}
\omega := \sum\limits_{i=1}^{\infty} 2^{-i} \delta_{(0,0)} \times \delta_{(2^{i/s},0)}
\end{align*}
and therefore
\begin{align*}
LW_s(\mu,\nu) & \geq \left[ \int\limits_{\M^2} d^s d\omega \right]^s = \left[ \sum\limits_{i=1}^{\infty} d\left( (0,0),(2^{i/s},0) \right)^s 2^{-i} \right]^s = \left[ \sum\limits_{i=1}^{\infty} \left( 2^{i/s} \right)^s 2^{-i} \right]^s = \left[ \sum\limits_{i=1}^{\infty} 1 \right]^s
\\
& = +\infty.
\end{align*}
\end{Ex}
However, Lorentz--Wasserstein distances between two compactly supported measures in globally hyperbolic spacetimes \emph{are} finite.
\begin{Prop}
Let $\M$ be a~globally hyperbolic spacetime, $s \in (0,1]$ and let $\mu,\nu \in \mathfrak{P}(\M)$ be compactly supported. Then, $LW_s(\mu,\nu) < +\infty$.
\end{Prop}
\begin{Proof}
If $\Pi_c(\mu,\nu) = \emptyset$, then trivially $LW_s(\mu,\nu) = 0 < +\infty$. Assume then that the~set of causal couplings between $\mu$ and $\nu$ is nonempty and take any $\omega \in LW_s(\mu,\nu)$. On the~strength of Proposition \ref{couplings}, $\omega(\supp \, \mu \times \supp \, \nu) = 1$. By assumption, the~set $\supp \, \mu \times \supp \, \nu \subseteq \M^2$ is compact. Moreover, by the~global hyperbolicity of $\M$, $d$ is a~continuous map and hence it is bounded on that compact set. Therefore,
\begin{align*}
& \int\limits_{\M^2} d(p,q)^s d\omega(p,q) = \int\limits_{\supp \, \mu \times \supp \, \nu} d(p,q)^s d\omega(p,q) + \underbrace{\int\limits_{\M^2 \setminus \left( \supp \, \mu \times \supp \, \nu \right)} d(p,q)^s d\omega(p,q)}_{= \, 0, \textnormal{ because the~domain of integration is } \omega\textnormal{-null}}
\\
& \leq \max\limits_{\substack{p \, \in \, \supp \, \mu \\ q \, \in \, \supp \, \nu}} d(p,q)^s \int\limits_{\supp \, \mu \times \supp \, \nu} d\omega = \left[ \max\limits_{\substack{p \, \in \, \supp \, \mu \\ q \, \in \, \supp \, \nu}} d(p,q) \right]^s
\end{align*}
\noindent
and so, by the~arbitrariness of $\omega$,
\begin{align*}
LW_s(\mu,\nu) \leq \max\limits_{\substack{p \, \in \, \supp \, \mu \\ q \, \in \, \supp \, \nu}} d(p,q) < +\infty.
\end{align*}
\end{Proof}

\section{Outlook}\label{sec:outlook}

Let us briefly summarise the~main results of the~paper. We proposed a~notion of a~causal relation between probability measures on a~given spacetime $\M$. To give sense to Definition \ref{causality_def_true} embedded in the~theory of optimal transport, we had to enter the~domain on the~verge of causality and measure theory. We believe that our paper paves the~way to this \textit{terra incognita}, which is worth exploring both from the~viewpoint of mathematical relativity, as well as possible applications in quantum physics.

On the~mathematical side, the~presented theory can be developed in various directions.

Firstly, one can try to lower the~causality conditions imposed on the~spacetime in the~theorems presented in Section \ref{sec:main}. In particular, it would be interesting to see whether the~defined relation on $\PP(\M)$ is a~partial order for every causal spacetime $\M$, or is the~assumption (\ref{propertyP}) in Theorem \ref{antisymmetry_partial} a~necessary one. If the~latter holds, one would obtain a~new rung of the~causal ladder between the~causal and distinguishing spacetimes.

A~second path of possible development is to investigate further the~notion of a~Lorentzian distance in the~space of probability measures on a~spacetime, and the~associated topological questions.  In Section \ref{sec:LW} we proposed a~notion of the~$s^\textnormal{th}$ Lorentz--Wasserstein distance, which is a~natural generalisation of the~Lorentzian distance between the~events on $\M$. However, in the~optimal transport theory there are other ways to measure distances between probability measures (see for instance \cite[p. 97]{Villani2008}). It is tempting to see how (if at all) these notions can be adapted to the~spacetime framework. This directly relates to the~issue of topology on $\PP(\M)$ and its interplay with the semi-Riemannian metric on $\M$.

Another potential direction of future studies, particularly interesting from the~viewpoint of applications, would be to generalise the~results of the~present paper to signed measures. This would allow to study causality of, both classical and quantum, charge (probability) densities on spacetimes.

The~applications of the~developed theory in classical and quantum physics will be discussed in details in a~forthcoming paper. Let us, however, make some remarks here.

Probability measures on space(time) arise in a~natural way in quantum theory from the~wave functions via the~`modulus square principle'. The~results of Hegerfeldt show that in a~generic quantum evolution driven by a~Hamiltonian bounded from below a~state initially localised in space immediately develops infinite tails. If a~quantum system is acausal in the~sense of Hegerfeldt, then it is so in the~sense of Definition \ref{causality_def_true}. Indeed, a~wave function is localised (that is of compact support) if and only if the~corresponding probability measure is so. Thus, if $\mu_0 \in \PP(\R^n)$ has compact support and $\mu_t \in \PP(\R^n)$ extends to infinity for any $t>0$, then $\delta_0 \times \mu_0 \npreceq \delta_t \times \mu_t$ as measures on the~$(n+1)$-dimensional Minkowski spacetime on the~strength of Proposition \ref{neccond}.

Note however, that Proposition \ref{neccond} provides only a~\emph{necessary} condition for a~causal relation to hold, and not a~\emph{sufficient} one. In \cite{Hegerfeldt1985}, Hegerfeldt has extended his theorem to initial states with exponentially bounded tails. He also suggested therein that a~similar phenomenon resulting in the~breakdown of causality should occur for states with powerlike decay. It thus indicates that acausality is a~property of the~quantum system and cannot be avoided by the~use of nonlocal states. Our Definition \ref{causality_def_true} opens the~door to check this conjecture in a~mathematically rigorous way.

It is sometimes argued (see for instance \cite{MinimalPacket,Barat2003}) that Hegerfeldt's theorem implies that localised quantum states do not exist in Nature. This conclusion is however challenged by the~results in \cite{ZitterQFT}, which suggest that there is no lower limit on the~localisation of the~electron. Moreover, the~fact that a~state is nonlocal does not necessarily cure the~causality violation. Indeed, imagine that one disposes of an~initial quantum state, localised or not, which undergoes an~acausal evolution, i.e. $\delta_0 \times \mu_0 \npreceq \delta_t \times \mu_t$ for any $t>0$. Then one could encode information in the~probability density of $\mu_0$ in some compact region $\K$ of space and transmit it to an~observer localised outside of $J^+(\{0\} \times \K)$, as follows from the~condition $4^\bullet$. Such a~method of signalling would have a~very low efficiency, but is \emph{a priori} possible -- see for instance the~discussion in \cite{Hegerfeldt2001} and other cited works by Hegerfeldt.

Finally, let us come back to the~original motivation of our preliminary Definition \ref{causality_def1}. As stressed at the~beginning of Section \ref{sec:main}, it was inspired by the~notion of `causality in the~space of states' coined in \cite{CQG2013}. The~partial order relation considered in \cite{CQG2013} is defined on the~space of states $S(\A)$ of a~$C^*$-algebra $\A$. If the~algebra $\A$ is commutative then, by Gelfand duality, there exists a~locally compact Hausdorff topological space $\M$, such that $\A \simeq C_0(\M)$. Then, the~Riesz--Markov representation theorem implies that $S(\A) \simeq \PP(\M)$. Hence, if $\M$ is a~causally simple spacetime, then the~two notions of `causality for Borel probability measures' and `causality in the~space of states' coincide.

The~concept of causality in the~space of states was explored \cite{SIGMA2014,JGP2015} in the~framework of `almost commutative spacetimes', i.e. for $C^*$-algebras of the~form $C_0(\M) \otimes \A_F$, with $\A_F$ being a~finite dimensional matrix algebra. However, the~study therein was limited only to special subclasses of all states, nevertheless yielding interesting results. The~theory put forward in the~present paper blazes a~trail to unravel the~complete causal structure of almost commutative spacetimes. Having in mind that almost commutative spacetimes are utilised to build models in particle physics \cite{WalterBook}, it is enticing to see whether the~extended causal structure imposes any restrictions on probabilities that could be checked experimentally.

\bibliographystyle{abbrv}
\bibliography{causality_bib}{}

\end{document}